\documentclass[aps,preprint,amssymb,12pt,floatfix]{revtex4}
\pdfoutput=1

\usepackage{subfigure}
\usepackage{graphicx}
\usepackage{lscape,graphicx}
\usepackage{rotating}
\usepackage{epstopdf}
\usepackage{color}
\usepackage{amsmath,amsthm}
\usepackage{wrapfig}
\usepackage{enumerate}
\usepackage{fancyhdr}
\usepackage{pdfpages}
\usepackage{hyperref}

\begin{document}

\title[Allostery of GPCRs]{Mapping the intramolecular signal transduction of G-protein coupled receptors}

\author{Yoonji Lee$^1$, Sun Choi$^{1,\ast}$, and Changbong Hyeon$^{2,}$}
\thanks{sunchoi@ewha.ac.kr; hyeoncb@kias.re.kr}
\affiliation{
$^1$National Leading Research Lab of Molecular Modeling \& Drug Design, College of Pharmacy, Graduate School of Pharmaceutical Sciences, and Global Top 5 Research Program, Ewha Womans University, Seoul 120-750, Korea \\
$^2$School of Computational Sciences, Korea Institute for Advanced Study, Seoul 130-722, Korea
}

\begin{abstract}
G-protein coupled receptors (GPCRs), a major gatekeeper of extracellular signals on plasma membrane, are unarguably one of the most important therapeutic targets. 
Given the recent discoveries of allosteric modulations, an allosteric wiring diagram of intramolecular signal transductions would be of great use to glean the mechanism of receptor regulation.
Here, by evaluating betweenness centrality ($C_B$) of each residue, we calculate maps of information flow in GPCRs and identify key residues for signal transductions and their pathways. Compared with preexisting approaches, the allosteric hotspots that our $C_B$-based analysis detects for A$_{2A}$ adenosine receptor (A$_{2A}$AR) and bovine rhodopsin are better correlated with biochemical data.
In particular, our analysis outperforms other methods in locating the rotameric microswitches, which are generally deemed critical for mediating orthosteric signaling in class A GPCRs. 
For A$_{2A}$AR, the inter-residue cross-correlation map, calculated using equilibrium structural ensemble from molecular dynamics simulations, reveals that strong signals of long-range transmembrane communications exist only in the agonist-bound state. 
A seemingly subtle variation in structure, found in different GPCR subtypes or imparted by agonist bindings or a point mutation at an allosteric site, can lead to a drastic difference in the map of signaling pathways and protein activity. 
The signaling map of GPCRs provides valuable insights into allosteric modulations as well as reliable identifications of orthosteric signaling pathways. 
\end{abstract}

\maketitle
\section*{Introduction}
G-protein coupled receptors (GPCRs), a major gatekeeper on the cell surface, mediate various physiological processes, such as vision, olfaction, cardiovascular function, and immune responses, which makes GPCRs one of the most important therapeutic targets \cite{Rosenbaum2009Nature}. Consisting of seven $\alpha$-helical transmembrane (TM) domains, extracellular and intracellular loops (ECLs and ICLs), GPCRs relay extracellular signals to the cytoplasmic domain and activate proteins associated with signal transduction pathways \cite{Rosenbaum2009Nature,Jensen2004EJPS}. The activity of GPCRs is highly selective to the type of extracellular signals \cite{wess1997FASEBJ}, and is sensitively modulated by point mutations \cite{Kjelsberg1992JBC,Bond06TPS}, the latter of which is closely related to the development of drug resistance as well \cite{Yun08PNAS,Yonath05ARBiochem}. 
Because agonist binding to orthosteric sites enables accommodation and activation of G-protein by regulating the conformational change in cytoplasmic domain \cite{Rosenbaum2009Nature}, developing antagonist or agonist drugs targeting at orthosteric sites has been a straightforward strategy of drug design. Such strategy, however, has often shown limited success due to the high sequence conservation among the members of a GPCR subfamily. Instead, there have been several reports on the efficacy of allosteric modulators in ion-channels \cite{Galzi94COSB} and other systems \cite{Changeux2012ARB} as well as GPCRs \cite{liu2012Science}, which highlights the role of allosteric sites in regulating the orthosteric signaling. 

Both orthosteric signaling and allosteric modulation are associated with long-range communications between two remote sites in the receptor structure. Such communications, which are altogether referred to as allostery, could be a consequence of a special balance of intramolecular forces formed in the network of the inter-residue contacts. Even though overall backbone topologies are similar between two different receptors belonging to the same subfamily, the sequence variation alters the connectivity map or local packing, which could be led to drastic differences in the allosteric signaling map. Despite long history of study on protein dynamics and even with the atomistic details of three-dimensional structure at hand, the structural basis of protein allostery still remains elusive and stubbornly resists revealing its microscopic underpinnings. While protein mutagenesis is a standard experimental method to identify key residues for protein function, the associated experiments are laborious and time-consuming. To overcome this experimental difficulty, there has been a growing interest in the use of theoretical/bioinformatics analysis. 
Careful statistical evaluations of multiple sequence alignment (MSA) of a protein family can be used to detect a set of evolutionally correlated residues \cite{Lockless99Science,Suel2002NSMB,Dima06ProtSci,morcos2011PNAS}. In addition, variants of normal mode analysis have been proposed to identify key residues that control ``functional" motion of enzymes and molecular motors \cite{Zheng05Structure,Zheng06PNAS,BaharMSB06,balabin2009PNAS}. 
Although a complementary use of these methods with molecular simulations would hold good promise to decipher the allosteric network of residues that are critical for the functional dynamics of proteins \cite{Lee11JACS}, a certain class of residues are still difficult to identify if the sequence conservation of the residues is too strong, or if the residues, deeply buried at the core regions of proteins, show only a minor conformational change along with the global conformational dynamics of protein. 
For instance, in class A GPCRs, 18 key residues buried in the TM region, called microswitches (or rotamer toggle switches) \cite{Rosenbaum2009Nature,nygaard2009TPS} belong to such a class (see below). 

Here, we propose a simple but powerful method to calculate the map of allosteric signal flow within protein structure, which identifies core allosteric sites including the above-mentioned microswitches. 
For a protein structure represented as a network of residues, we used a measure in the network theory called ``betweenness centrality" ($C_B$) \cite{freeman1979SocialNetworks,newman2005SocialNetworks} to evaluate the importance of each residue from the perspective of the flow of information (see Materials and Methods). 
By adopting A$_{2A}$ adenosine receptor (A$_{2A}$AR) and other GPCRs as model systems, we decide residues important for the allosteric signaling and pathways of signal flow.
The comparison of the results from our analysis with those from other methods shows that the $C_B$-based network analysis of protein structure is much simpler, but is more reliable in identifying the allosteric hotspots that includes microswitches. Furthermore, allosteric hotspots are identified from another analysis adapting the concept of network vulnerability \cite{Jeong01Nature,delSol2006MSB}; and explicit calculation of individual multiple pathways linking the clusters of long-range correlated residues across transmembrane shows that majority of paths pass through the hotspot residues we predicted. 
The predictions from the $C_B$-based network analysis of protein structure should be of great use not only to complement mutagenesis study but also to elucidate the origin of subtype selectivity as well as the activation and regulation mechanisms of GPCRs. 

\noindent\section*{Materials and Methods}

\noindent{\bf Quantification of sequence conservation.}
For a given multiple sequence alignment (MSA) of a protein family, the following statistical free energy-like function scaled by an arbitrary energy scale $k_BT^*$ quantifies the extent of sequence conservation \cite{Dima06ProtSci,Lee11JACS}: 
\begin{align}
\Delta G_i/k_BT^*=\sqrt{\frac{1}{C_i}\sum_{\alpha=1}^{20}\left[p_i^{\alpha}\log{\frac{p_i^{\alpha}}{p_{\alpha}}}\right]^2},
\label{eqn:seq}
\end{align}
where $C_i$ is the number of amino acid types at position $i$ along the sequence, $\alpha$ denotes amino acid species, $p_i^{\alpha}$ is the frequency of an amino acid $\alpha$ at the position $i$, and $p_{\alpha}$ is the frequency of an amino acid $\alpha$ in the full MSA, which serves as the background frequency. 
Note that the quantity ``$S=\sum_{\alpha=1}^{20}p_i^{\alpha}\log{(p_i^{\alpha}/p_{\alpha})}$" is the relative entropy; $S=0$ if $p_i^{\alpha}$ is no different than $p_{\alpha}$ for all $\alpha$. 
The larger the value of $\Delta G_i$, the sequence at the position $i$ is better conserved. 
In this paper, we computed $\Delta G\mathrm{(GPCR)}/k_BT^*$ and $\Delta G\mathrm{(AR)}/k_BT^*$, 
each of which is evaluated using different MSA.  
To obtain the MSAs of AR and GPCR families, we collected the sequences of adenosine receptor family (219 sequences) and class A GPCR family (26,655 sequences) from UniProtKB and Pfam database, respectively. 
After filtering the redundancy, 208 sequences and 24,507 sequences were remained for AR and GPCR family. 
For GPCR, sequence clustering was performed with 40\% identity to reduce the sequence space size, and 2,471 sequences were obtained. Based on these sequences, the multiple sequence alignment (MSA) was produced using the log-expectation (MUSCLE) program \cite{edgar2004muscle}. \\

\begin{figure}[!ht]
\begin{center}
\includegraphics[width=6in]{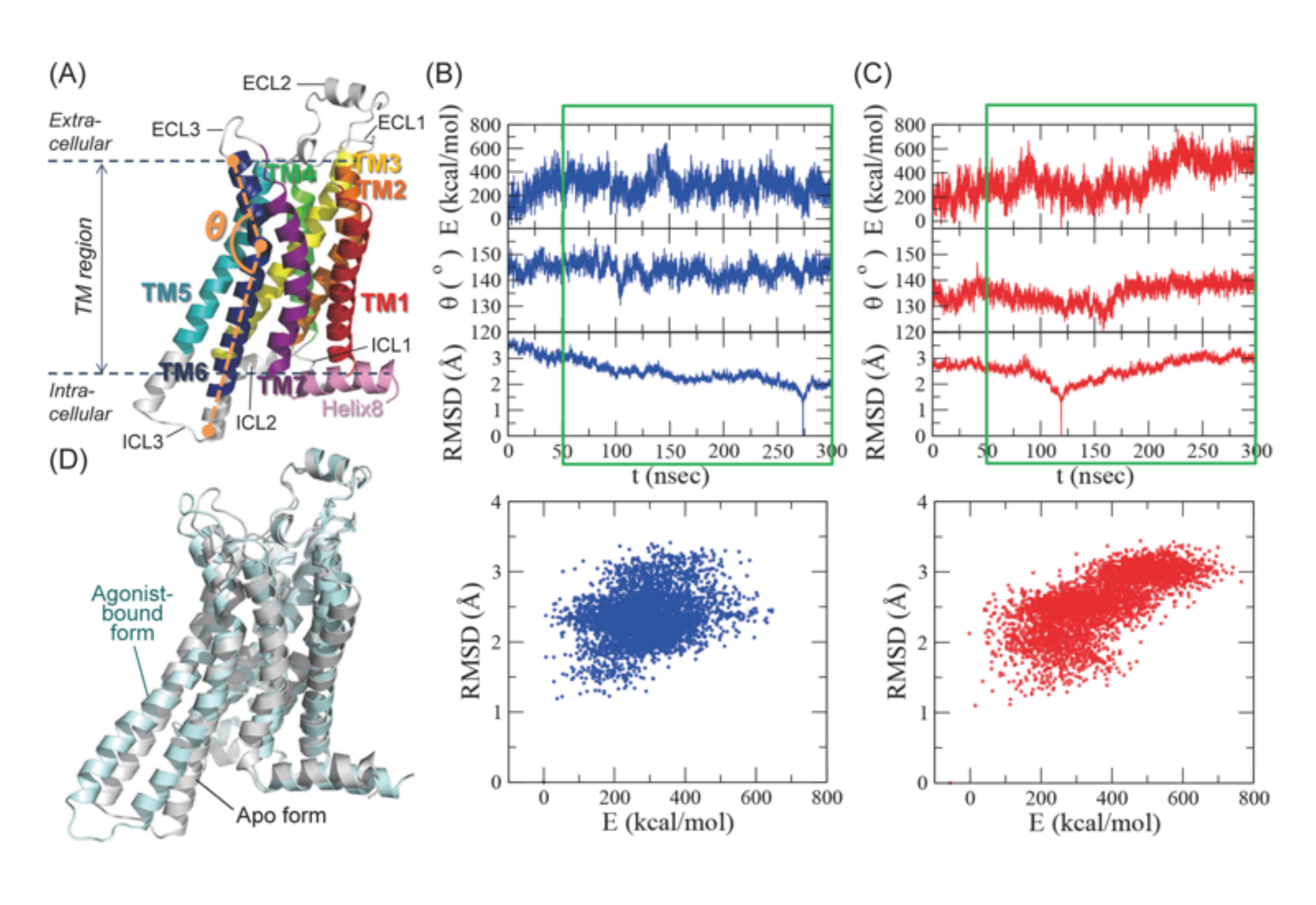}
\end{center}
\caption{
{\bf Structure and dynamics of A$_{2A}$ adenosine receptor.}  
(A) Seven TM helices and the intra- and extracellular loops. (B, C) Total conformational energy, TM6 tilt angle ($\theta$) measured between three points defined along the center of helix using three group of residues (255-258, 244-247, 219-222), and RMSD in reference to the minimum energy structure from the MD trajectories of the (B) apo and (C) agonist-bound forms. Analysis was carried out for the boxed time interval, which excludes the first 50 nsec trajectories. (D) The minimum energy structures of the A$_{2A}$ARs in the apo and agonist-bound forms are overlaid to show that the most significant difference between the two forms is in the cytoplasmic region of TM5-ICL3-TM6. 
}
\label{MD}
\end{figure}

\noindent{\bf Generating the minimal energy structures and conformational ensemble of the human A$_{2A}$ adenosine receptor.}
The conformational flexibility of GPCRs makes it difficult to obtain high-resolution X-ray crystal structures, particularly, in the active state. Although several X-ray crystal structures of the A$_{2A}$AR are determined in their antagonist or agonist-bound forms \cite{xu2011Science,jaakola2008Science}, structural information in the apo form or fully active state is not yet available \cite{deupi2011COSB}. To prepare the human A$_{2A}$AR models (residues from I3 to Q310) including all the loop regions, homology modeling was performed using MODELER program implemented in Discovery Studio v.3.1. We used the structures with PDB IDs, 3EML \cite{jaakola2008Science} and 3QAK \cite{xu2011Science} as templates for the apo and agonist-bound forms, respectively, and 2YDV \cite{lebon2011Nature} and 3PWH to generate models for the loop regions that were not determined in 3EML and 3QAK. Conserved disulfide bridges, C71-C159, C74-C146, C77-C166 and C259-C262, were retained, and the agonist ligand was inserted to the agonist-bound-form model. The models were optimized with simulated annealing and selected based on the DOPE score. The final homology structures were obtained under GBSW implicit solvent hamiltonian by using conjugate gradient method. To generate the minimum energy structures and thermal structural ensemble of the A$_{2A}$ARs, we performed molecular dynamics simulation for 300 nsec with the NAMD v2.8 package using the CHARMM22/CMAP force field \cite{Buck2006BiophysJ}. To construct an explicit membrane system, the TM region of the A$_{2A}$AR was predicted based on the Orientations of Proteins in Membranes (OPM) database and the palmitoyloleoylphosphatidylcholine (POPC) membrane was placed around the TM region of the receptor. Then, the receptor in membrane system was solvated with the explicit water molecules and ionized with 150 mM KCl. The whole system was energy minimized in the order of lipid membrane, waters, and the entire molecules, followed by the heating, equilibration and production runs for 300 nsec under NPT ensemble. The trajectories of production run were monitored in terms of total conformational energy, tilt angle of TM6, and root mean square deviation relative to the initial ($t = 0$) structure. In accord with the common notion for GPCR dynamics, the tilt angle of TM6 varied between 135$^{\circ}$ and 150$^{\circ}$ for the apo form, and between 120$^{\circ}$ and 145$^{\circ}$ for agonist-bound form (Figure \ref{MD}B). Finally, the minimal energy conformations from the simulated trajectories were obtained for the apo and agonist-bound forms. Our minimal energy conformation for the agonist-bound form has a tilt angle 133$^{\circ}$ in TM6, whereas the agonist-bound crystal structure 3QAK has a tilt angle of 142$^{\circ}$. \\

\noindent{\bf Statistical assessment of a prediction method.}
Since there are 18 microswitches in class A GPCRs, the probability ($p_m$) of correctly identifying at least one microswitch out of 308 residues of GPCRs is given by $p_m=\frac{18}{308}\approx 0.06$. 
Then the expectation value of identifying microswitches by randomly drawing $n$ residues is $\langle n\rangle_{rand}=n\times p_m$. 
Thus, if $N_m$ microswitches are identified with a certain method, one can evaluate prediction efficiency of the method by calculating the ratio between $N_m$ and $\langle n\rangle_{rand}$, i.e., 
\begin{equation}
\varphi_m=\frac{N_m}{\langle n\rangle_{rand}}.  
\label{eqn:varphi}
\end{equation}
\\

\noindent{\bf Construction of the residue interaction network.}
We constructed the residue interaction network by representing each amino acid residue as a single node. To take into account the effect of side chain, we considered two coarse-grained centers per residue, i.e., C$\alpha$ carbon for backbone and a farthest heavy atom from C$\alpha$ for the side chain. By doing so, we included the cases of backbone-backbone, backbone-side chain, side chain-side chain contacts. In our network model, a link was established between two C$\alpha$-carbons when any pair of backbone and side chain of two residues is less than 7 \AA\ \cite{da2009protein}, thus the side chains are implicit in the network. \\

\noindent{\bf Network centralities.}
Simplifying architecture of complex system into a network (graph), which is represented with ``nodes" (vertices) and ``links" (edges), can be used as a powerful tool to extract key properties of the system topology and its components \cite{Watts98Nature}. Originally devised for analyzing social phenomena and later actively extended to reveal hub proteins central to the cellular, regulatory, metabolic networks as well as network property of each organism \cite{albert2000Nature,jeong2000Nature,balazsi2005PNAS}, network analysis can be carried out for studying protein structures as well. In the last decade, much attention has been paid in this direction. As a general statistical property of protein structure networks, networks of folded proteins display small-worldness, but are not scale-free \cite{vendruscolo2002PRE,atilgan2004BJ,greene2003JMB}. By quantifying key network properties for monomeric protein structures, one can address issues such as the plasticity of protein structures, folding of protein domains, and identify key residues along the folding pathways \cite{amitai2004JMB,bagler2007Bioinfomatics,brinda2005BJ,dokholyan2002PNAS}. In fact, the network analysis of protein structures can be extended further to identify key residues for allostery and their wiring diagram.
Several studies have recently been carried out to address the microscopic mechanism of protein allostery by applying the strategies of network or community analysis in conjunction with molecular dynamics simulation on model systems \cite{Sethi09PNAS,yao2013BJ} including GPCRs \cite{miao2013PNAS}.
To address the issue of allostery, we utilized the betweenness centrality, one of the most fundamental concepts in network analysis explained below, in identifying allosteric hotspots by surmizing that allosteric hotspots are the mediators of information flow in a network topology of a given protein structure.

\begin{figure}[!ht]
\begin{center}
\includegraphics[width=4in]{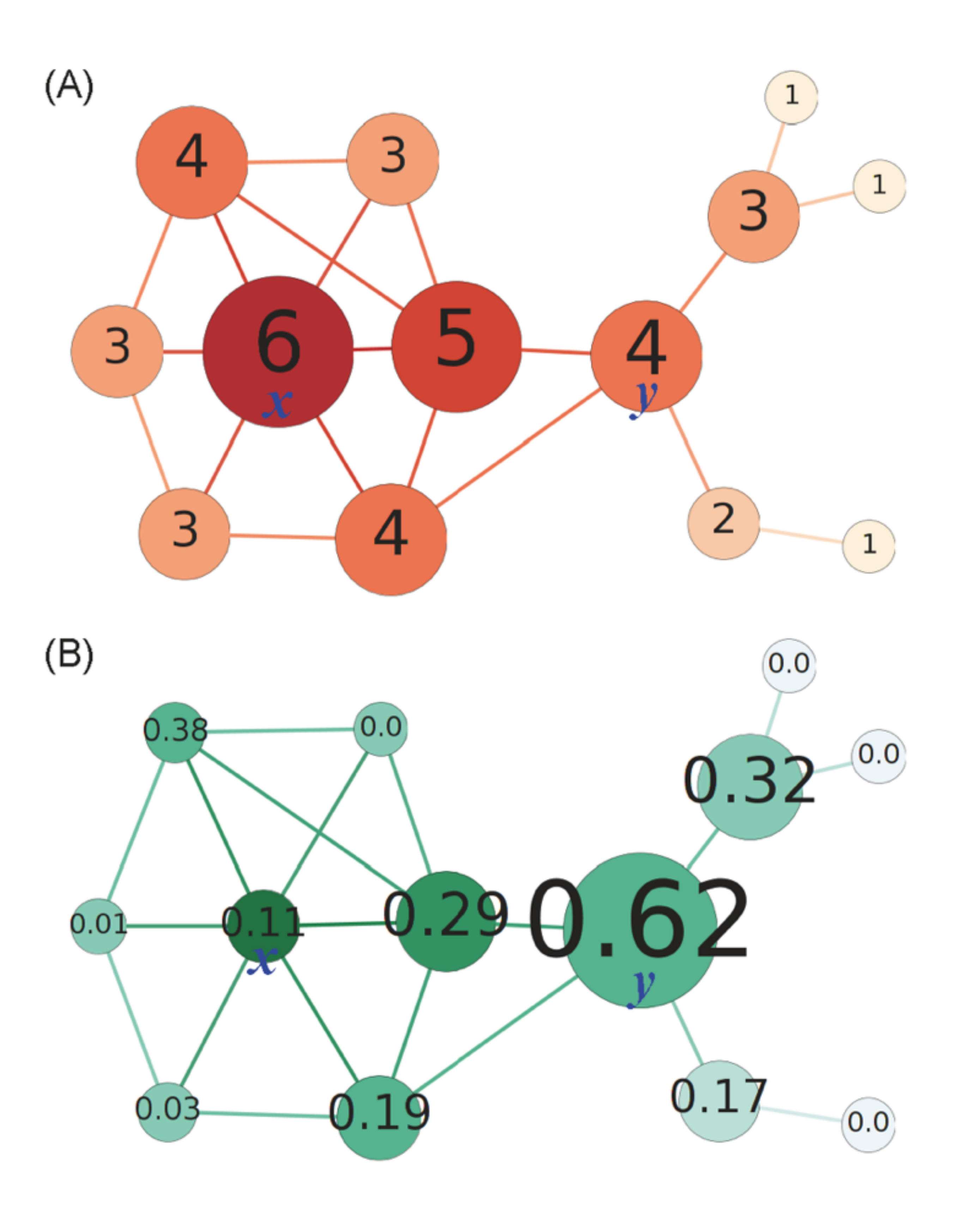}
\end{center}
\caption{
{\bf An example of graph showing the difference between the degree and betweenness centralities.}  
(A) Degree centrality (B) Betweenness centrality. 
The calculated centrality value is marked in each node.}
\label{centrality}
\end{figure}

Here the definitions are given for the three representative types of centrality for a node in a network: (1) The degree centrality $C_D(v)$ measures the number of edges linked to a node $v$.
\begin{equation}
C_D(v)=deg(v). 
\end{equation}
Note that $C_D(v)$ is identical to the number of contacts with its neighboring residues. (2) The closeness centrality $C_C(v)$, an inverse of mean geodesic distance (shortest path length) from all other nodes to the node $v$, measures how fast a signal from the node $v$ can be transmitted to other nodes. 
\begin{equation}
C_C(v)=\left(\sum_{i=1}^N{d(i,v)}/(N-1)\right)^{-1}, 
\end{equation}
where $d(i,v)$ is the minimal number of edges that bridge the nodes $i$ and $v$.
For a given network topology, $d(i,j)$ can be calculated by using Dijkstra's algorithm \cite{Dijkstra59NM}.   
(3) The betweenness centrality is the measure of the extent to which a node has control over transmission of information between the nodes in the network, which is defined as \cite{newman2005SocialNetworks}: 
\begin{equation}
C_{B}(v)=\frac{2}{(N-1)(N-2)}\sum_{s=1}^{N-1}{\sum_{t=s+1}^N{\frac{\sigma_{st}(v)}{\sigma_{st}}}},
\label{eqn:CB}
\end{equation}
where $s\neq t\neq v$. In the above definition, $\sigma_{st}$ is the number of shortest paths linking the nodes $s$ and $t$, and $\sigma_{st}(v)$ is the number of shortest paths linking the nodes $s$ and $t$ via the node $v$ \cite{newman2005SocialNetworks}. To calculate $C_B(v)$, we used Brandes algorithm \cite{brandes2001JMS}, which can reduce the computational cost of Eq. \ref{eqn:CB} substantially. The factor $\frac{(N-1)(N-2)}{2}$ is the normalization constant. 
The significance of betweenness centrality is succinctly illustrated in Figure \ref{centrality} using a graph where both $C_D$ and $C_B$ values are computed at each node. The node $x$ has a greater connectivity ($C_D=6$) to other nodes but its removal from the network does not destroy the communication among other nodes. 
In contrast the node $y$ has less connectivity ($C_D=4$) than $x$; yet upon removal of $y$ the whole graph would be split into three pieces. In the light of communication or the flow of information the node $y$ is the most critical. 
Note that $y$ has the highest $C_B$ value among the whole nodes.
Although a few studies \cite{delSol2006MSB,Atilgan2007BJ,park2011BMCbioinfo,daily2009PLoSCB} might appear similar in spirit to our work in that they also use centrality measures and shortest paths to decipher the allostery, it should be noted that different centrality measure has different assessment of each node. 
The \emph{betweenness centrality}, which evaluates the importance of each node based on the amount of traffic or the amount of inter-node communication, is one of the most ideal measures to identify allosteric hotspots for a given protein structure. 
\\

\section*{Results and Discussion}
\begin{figure}[!ht]
\begin{center}
\includegraphics[width=4.3in]{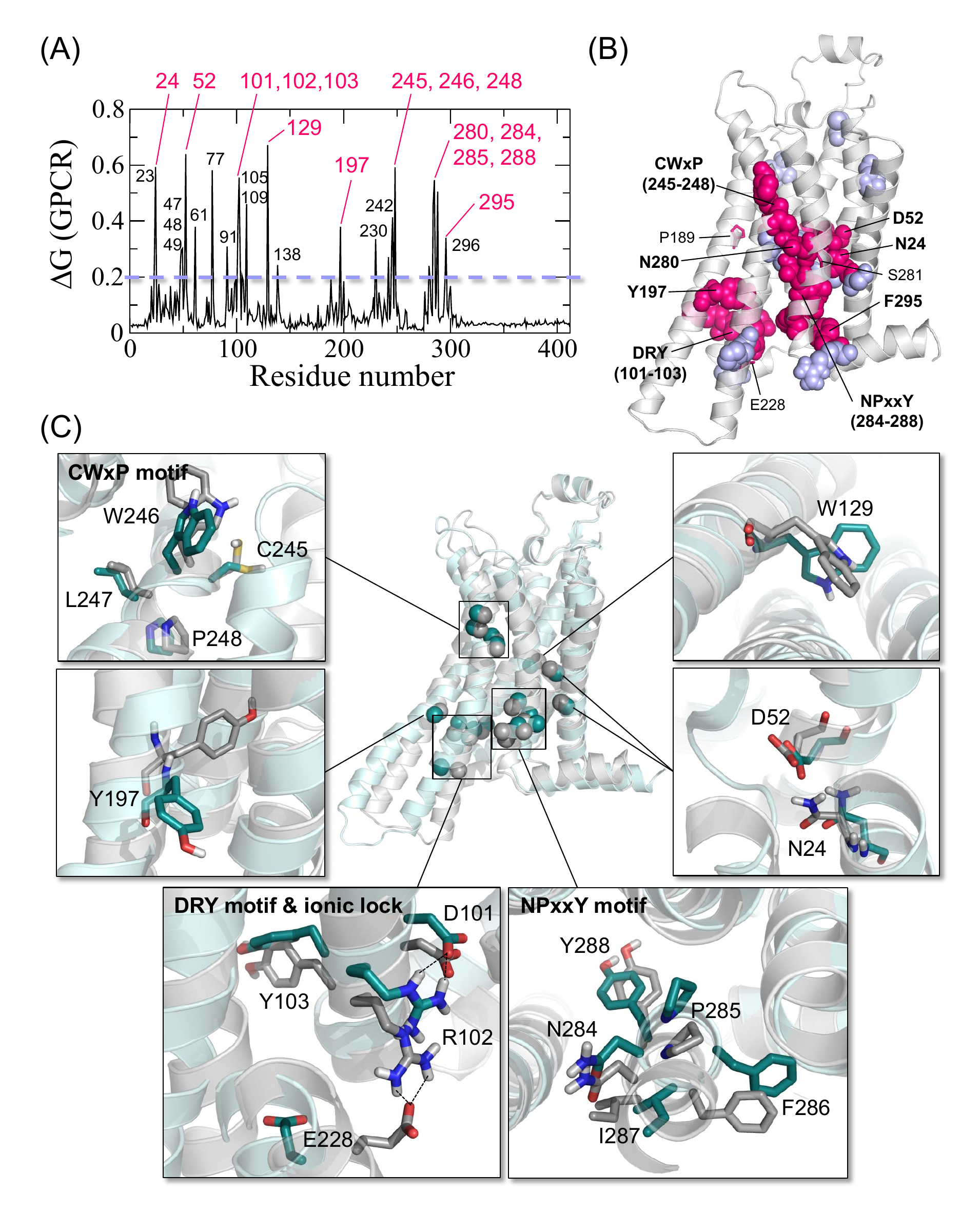}
\end{center}
\caption{{\bf Microswitches in GPCRs.} 
(A) Sequence conservation free energy ($\Delta G/k_BT^*$) computed for the class A GPCR family. 
The specification of GPCR inside the parenthesis of $\Delta G\mathrm{(GPCR)}$ in Figure \ref{microswitch} indicates that $\Delta G/k_BT^*$ value was calculated for a multiple sequence alignment for the class A GPCR family.  
The residues with $\Delta G\geq  0.2$ are annotated; and among them 15 residues identified as microswitches in literatures are highlighted in magenta. 
(B) Residues of $\Delta G/k_BT^*\geq 0.2$ are depicted with spheres on A$_{2A}$AR structure. 
Among them, microswitches are colored in magenta, and others with $\Delta G/k_BT^*>0.2$ are in light-blue. 
The residues P189, E228, and S281, that are proposed as microswitches in literatures but have $\Delta G/k_BT^*$ less than 0.2, are depicted using stick representation. 
(C) Conformational changes of the key structural motifs and microswitch residues are depicted using the minimal energy structures of apo and agonist-bound forms obtained from our MD simulations. 
It is proposed that the rotameric transitions of microswitches are critical for the intra-molecular signal transmission of GPCRs.} 
\label{microswitch}
\end{figure}

\noindent{\bf Microswitches: benchmarks for prediction tools on GPCR allostery.}
The activation mechanism of receptors belonging to class A GPCRs, which include adenosine, $\beta_{1,2}$-adrenergic, rhodopsin, chemokine, dopamine, histamine receptors, is believed to be accompanied by a global rearrangement of TM helices that helps accommodate the binding of G-proteins. 
In particular, the newly resolved X-ray crystal structure of the active form of $\beta_2$-adrenergic receptor complexed with heterotrimeric G-protein \cite{rasmussen2011Nature} has lent strong support on such proposal by clearly demonstrating that 10$^o$ outward tilt of the intracellular part of TM6 helix is essential for the full activation of the receptor. 
For the class A GPCRs, it has been suggested that the activation mechanism is regulated by 18 microswitches (N24, D52, D101, R102, Y103, W129, P189, Y197, E228, C245, W246, P248, N280, S281, N284, P285, Y288, F295) \cite{katritch2013ARPT,nygaard2009TPS}, which consist of DRY (D101$^{3.49}$, R102$^{3.50}$, and Y103$^{3.51}$ in TM3), CWxP (C245$^{6.47}$, W246$^{6.48}$, and P248$^{6.50}$ in TM6), and NPxxY (N284$^{7.49}$, P285$^{7.50}$, and Y288$^{7.53}$ in TM7) motifs \cite{hofmann2009Hofmann,suwa2011Pharmaceuticals} (where `x' stands for any amino acid residue and the numbers in the superscript of residues are based on the Ballesteros Weinstein numbering system \cite{ballesteros1995MethodsNeurosci}), and others. 
Historically these residues were first identified either by evaluating the sequence conservation among the class A GPCR family or by comparing the two structures of GPCR subtype in different states; and the functional importance of the selected residues was subsequently confirmed by mutagenesis studies \cite{Rosenbaum2009Nature, nygaard2009TPS}. 
Thus, a receptor belonging to the class A GPCRs is expected to utilize many of these 18 microswitches for allosteric signaling. 
Although one should still be mindful of the fact that the functional role of these microswitches have not been verified for all the GPCR subtypes, the 18 microswitches can be used as benchmark residues to assess the performance of a prediction tool on allosteric hotspots in GPCRs (see Materials and Methods). 
The extent of sequence conservation in class A GPCRs, quantified by evaluating the sequence conservation free energy $\Delta G$ (Eq.\ref{eqn:seq}) indicates that 15 out of 18 microswitches (except for P189, S281, and E228) satisfy $\Delta G\mathrm{(GPCR)}/k_BT^*\geq 0.2$; thus highly conserved (Figures \ref{microswitch}A and \ref{microswitch}B).  
Figure \ref{microswitch}C visualizes how rotameric transition is made from the inactive to active state and highlights the difference in the orientation of the side chain in some of the microswitches by contrasting the apo and agonist-bound states.
\\

\begin{figure}[!ht]
\begin{center}
\includegraphics[width=3.5in]{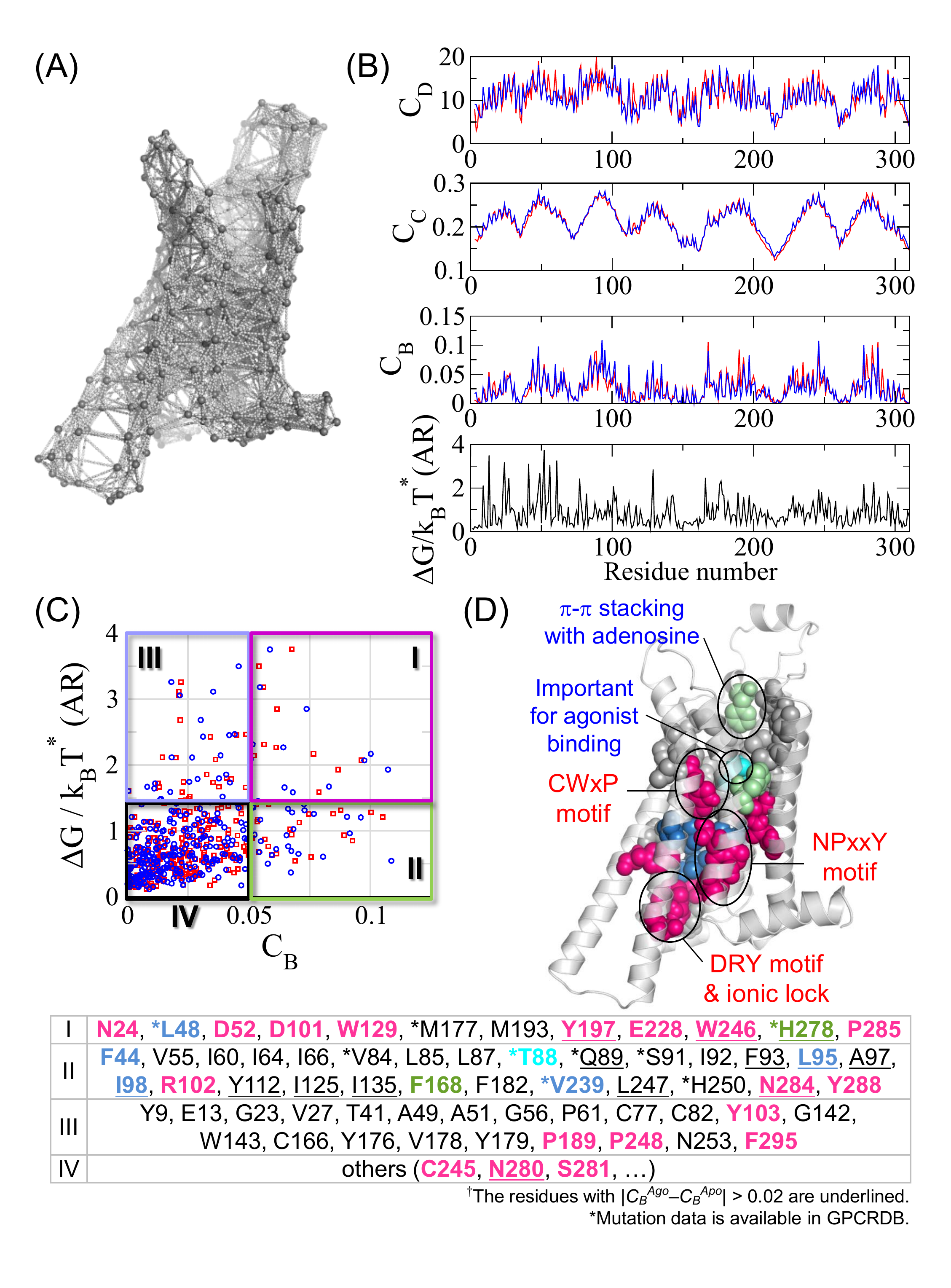}
\end{center}
\caption{
{\bf Network centrality analysis of A$_{2A}$ adenosine receptor.}  
(A) Network representation of the A$_{2A}$AR in apo form built by taking into account the presence of side chain in each residue (see Materials and Methods).   
(B) Degree ($C_{D}$), closeness ($C_{C}$), betweenness centralities ($C_{B}$) for each residue of the A$_{2A}$AR (blue: the apo form; red: the agonist-bound form) and the sequence conservation free energy ($\Delta G/k_BT^*$) calculated for AR family. 
(C) Scatter plot of ($\Delta G/k_BT^*$, C$_{B}$) (blue: apo form; red: agonist-bound form). 
Based on $C_{B}(i)$ and $\Delta G_i/k_BT^*$ values, the residues of A$_{2A}$AR were categorized into four groups from I to IV.  
The residues with high $C_{B} (\geq 0.05)$ (group I and II) and with high sequence conservation ($\Delta G/k_BT^* \geq 1.5)$ in AR family (group I and III) are depicted on the apo structure of the A$_{2A}$AR in Figures S2A and S2B, respectively. 
(D) Among the residues that belong to the groups I and II with $C^{Apo}_{B}\geq $ 0.05, 
key residues confirmed from the previous biochemical studies for class A GPCRs are marked on the A$_{2A}$AR structure using different colors (magenta for the microswitches (see Figure S3 for the top and bottom views): cyan for the residue important for agonistic binding; pale green for the residues important for ligand binding;  blue for the residues in hydrophobic barrier); the underlined residues satisfy the condition $|C_B^{Apo}-C_B^{Ago}|\geq 0.02$; and the residues marked with asterisks are those whose mutation data is available in GPCRDB for A$_{2A}$AR.}
\label{CB}
\end{figure}

\noindent{\bf The allosteric hotspots of A$_{2A}$ adenosine receptor mediate the flow of information.}
As a tool for studying protein allostery, the network centrality, a measure that quantifies the degree of centralization of a node in network theories, can be employed to unravel the hotspot residues of a given protein network. Among the popular centrality measures in network theories \cite{freeman1979SocialNetworks} (degree ($C_{D}$), closeness ($C_{C}$), and betweenness ($C_{B}$) centralities, whose definitions are given in Materials and Methods), the betweenness centrality, $C_{B}(v)$, evaluates the extent to which the node $v$ has control over the information flow in the network \cite{newman2005SocialNetworks}. 
Conceptually, it could be argued that a node of high $C_B$ value is the spot mediating the allosteric signal flow (Materials and Methods). 
By using the minimum energy structures form obtained from MD simulations we constructed the residue interaction network for the apo and agonist-bound states of the A$_{2A}$AR by taking into account the presence of side chains (Figure \ref{CB}A, see Materials and Methods), and calculated $C_{D}(v)$, $C_{C}(v)$, and $C_{B}(v)$ (Figure \ref{CB}B). 
The overall correlations between the different centrality measures are not that strong (correlation coefficient $=0.66-0.72$) (Figure S1); thus 
a residue with high $C_{D}$ (or equivalently with a large number of contacts) or $C_{C}$ does not necessarily retain a high $C_{B}$ value. 
Among the three network centralities, $C_{B}$ exhibits the highest selectivity (Figure \ref{CB}B). 
As depicted on the A$_{2A}$AR structure, the residues with $C_{B} \geq 0.05$ (the top 10 \% of the $C_B$-distribution), which are deemed important for controlling information flow from the definition of $C_B$, are distributed contiguously, bridging the extracellular (EC) and intracellular (IC) parts of TM helices (Figure \ref{CB}D).

The 39 allosteric hotspots of A$_{2A}$AR predicted using $C_{B} \geq 0.05$ (the residues are listed in the groups I and II in Figure \ref{CB}D) include many important residues suggested from biochemical studies for class A GPCRs in general and A$_{2A}$AR in particular. 
Among the 18 residues suggested as general microswitches for class A GPCRs, 11 of them (N24, D52, D101, R102, W129, Y197, E228, W246, N284, P285, Y288) are identified by the simple condition of $C_{B} \geq 0.05$.
Since the probability of correctly identifying at least a single microswitch from random drawing is $p_m=18/308\approx 0.06$, the expectation value of identifying microswitches by selecting 39 residues is $39\times p_m\approx 2.3$ (see Materials and Methods). 
Given that we identified as many as 11 residues the performance of our $C_B$-based analysis should be considered significant.  
Among the residues identified by this condition other than microswitches, 
F44$^{2.42}$, L48$^{2.46}$, L95$^{3.43}$, I98$^{3.46}$, and V239$^{6.41}$ (blue in Figure \ref{CB}D) compose a region called the hydrophobic barrier that separates CWxP and NPxxY motifs from DRY motif \cite{trzaskowski2012CMC}; 
F168 in ECL2, H278$^{7.43}$, and T88$^{3.36}$ (green in Figure \ref{CB}D) are the residues known to be important for ligand binding in AR family \cite{xu2011Science}. 
F168 can potentially interact with adenine ring of nucleoside ligands via $\pi$-$\pi$ stacking.  
T88$^{3.36}$ in the TM3 helix that can form a hydrogen bonding with an agonist is important for sensing the agonist binding and transmitting signals to the intracellular G-protein binding site \cite{xu2011Science,jacobson2005ChemBiol,kim2003JMC}; 
L48, M177, V84, T88, Q89, S91, H250 (marked with asterisks in Figure \ref{CB}D) are also found essential for receptor function of A$_{2A}$AR according to the mutation data in GPCRDB \cite{Horn2003NAR}; 
Lastly the residues identified by $C_B\geq 0.05$ but not commented above to have any overlap with the previous biochemical studies (M193, V55, I60, I64, I66, L85, L87, I92, F93, A97, Y112, I125, I135, F182, L247) could be regarded as candidate residues for allosteric hotspots of A$_{2A}$AR that our $C_B$-based analysis predicts, which are amenable to further experimental study.

In conjunction with $C_B$ value, the extent of sequence conservation in each residue, $\Delta G\mathrm{(AR)}/k_BT^*$ (Eq.\ref{eqn:seq}), based on the multiple sequence alignment of adenosine receptor subfamily, could be useful for the purpose of our analysis. 
Here, it should be noted that $\Delta G\mathrm{(AR)}$ is different from $\Delta G\mathrm{(GPCR)}$ in Figure \ref{microswitch}A.  
$\Delta G(AR)$ is calculated by restricting the MSA to the subfamily of adenosine receptors while $\Delta G(GPCR)$ is calculated using the entire MSA for class A GPCRs.  
Partitioning the residues into four different groups based on the $\Delta G/k_BT^*$ and $C_B$ scores (Figure \ref{CB}C), 
i.e., $C_B\geq 0.05$, $\Delta G/k_BT^*\geq1.5$ for group I;
$C_B\geq 0.05$, $\Delta G/k_BT^*<1.5$ for group II;
$C_B < 0.05$, $\Delta G/k_BT^*\geq 1.5$ for group III;
$C_B < 0.05$, $\Delta G/k_BT^*<1.5$ for group IV,   
we make a few points below. 

\begin{figure}[!ht]
\begin{center}
\includegraphics[width=6in]{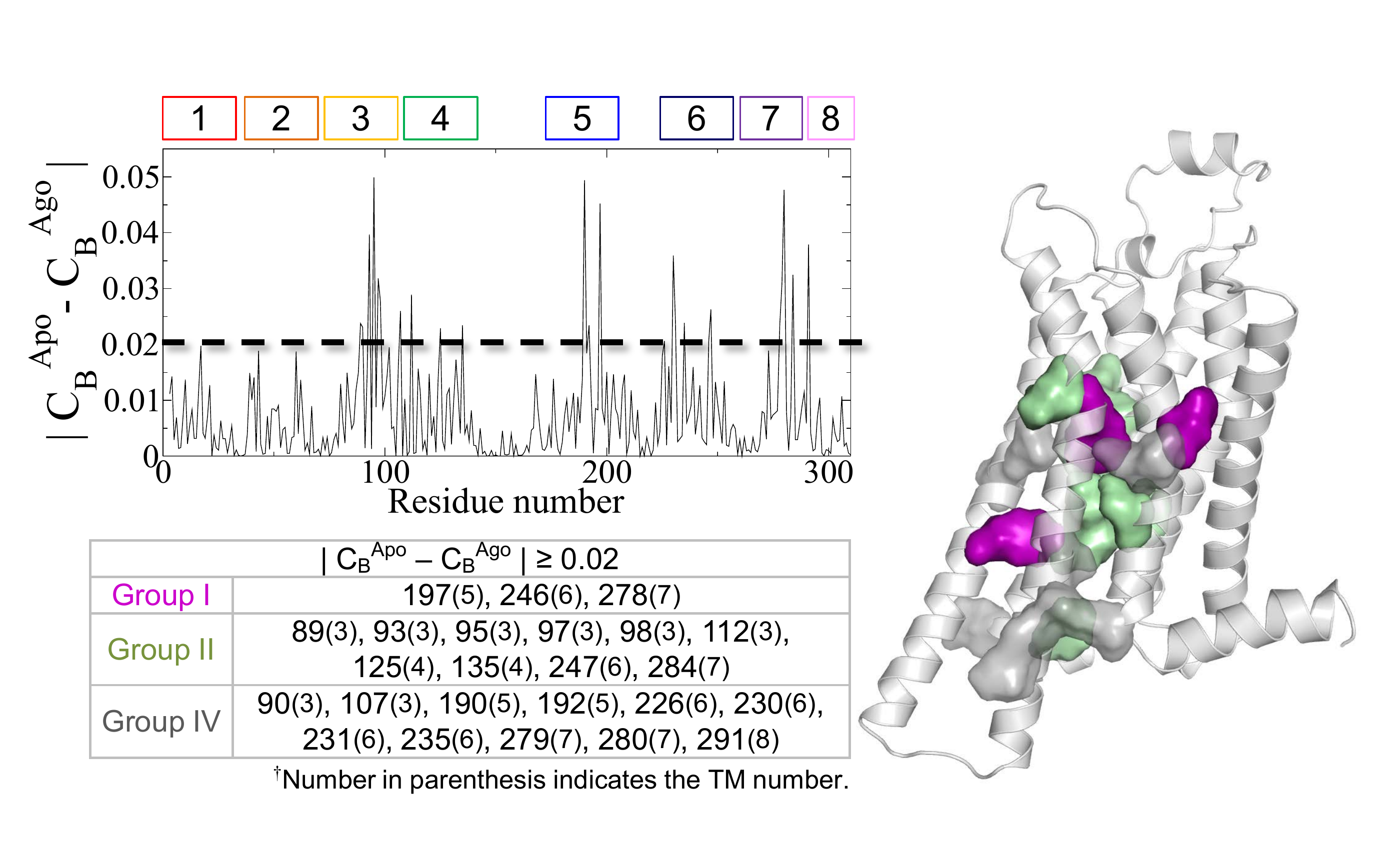}
\end{center}
\caption{{\bf Difference of $C_{B}$ values calculated for apo and agonist-bound structures.} Residues with $|C_B^{Ago}-C_B^{Apo}|\geq 0.02$, contributed from TM3, 5, 6, and 7, are depicted with magenta for group I, green for group II, grey for group IV, and their indices are listed on the table.
}
\label{CB_diff}
\end{figure}

\begin{figure}[!ht]
\begin{center}
\includegraphics[width=6in]{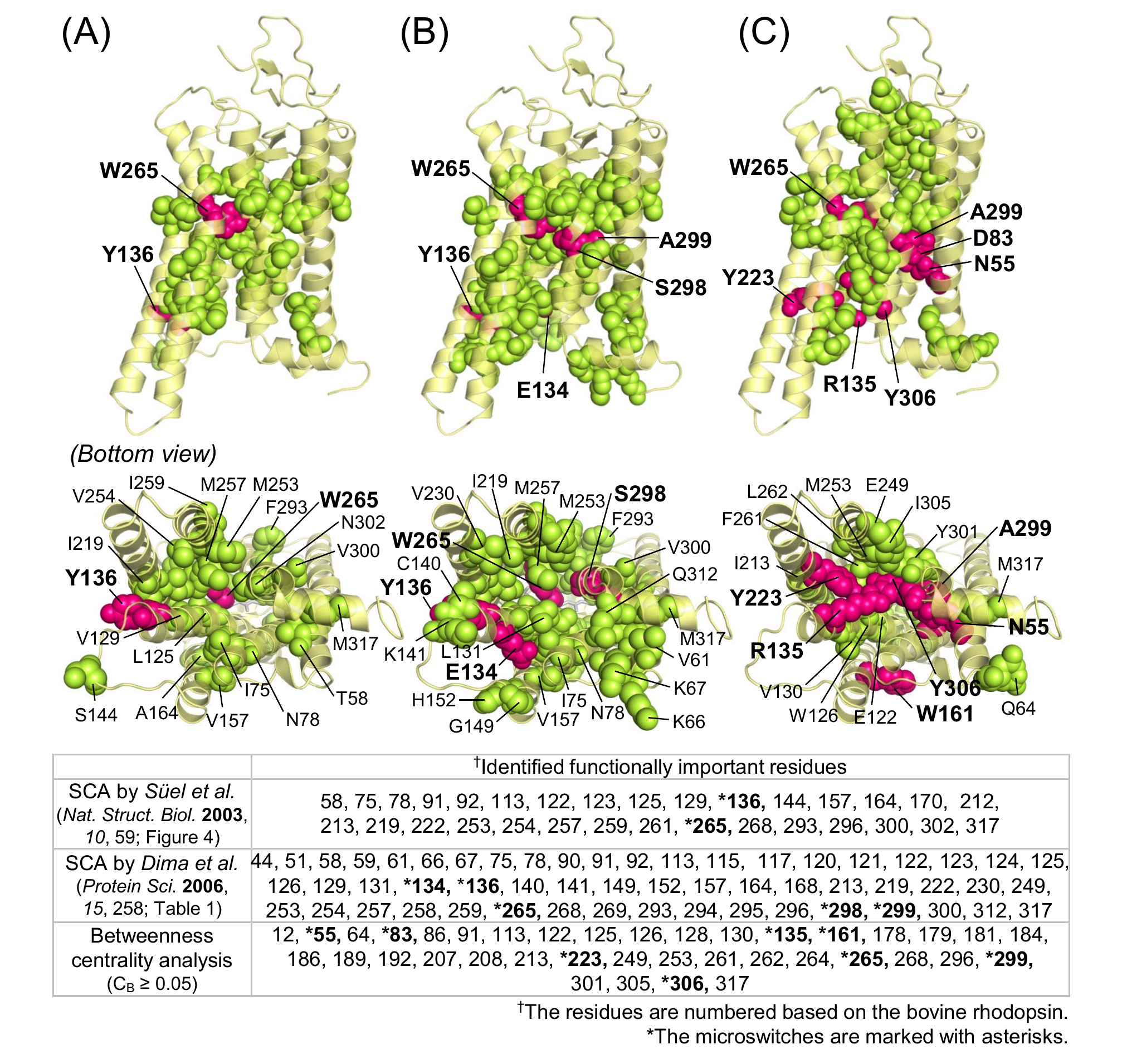}
\end{center}
\caption{
{\bf Comparison between the allosteric hotspots for rhodopsin predicted by SCA and $C_B$-based analysis.}  
Hotspots identified from SCA by (A) Suel \emph{et al.} \cite{Suel2002NSMB}, and (B) Dima \emph{et al.} \cite{Dima06ProtSci}, and (C) from our network analysis based on the residues with C$_{B} \geq 0.05$. Each method detected (A) 2 (B) 5 (C) 8 microswitches.} 
\label{SCA}
\end{figure}

\begin{enumerate}[(a)]

\item First, the definitions of $C_B$ and $\Delta G/k_BT^*$ are totally independent from each other. 
Evident from the scatter plot shown in Figure \ref{CB}C, no clear correlation is found between $C_B$ and $\Delta G\mathrm{(AR)}/k_BT^*$.   
Yet, the commonly identified residues with the conditions of high $C_B$ ($\geq 0.05$) and high $\Delta G/k_BT^*$ ($\geq 1.5$), namely the group I residues contains as many as 8 microswitches and 4 other hotspot residues. 
The group I residues ($C_{B} \geq 0.05$, $\Delta G/k_BT^* \geq 1.5$) are clustered at the core region of TMs (magenta region in Figures S2A and S2B), the contiguous surface of which is known to form hydrogen-bond network with the conserved polar residues and structural water molecules \cite{nygaard2009TPS,angel2009PNAS}. 
Of particular note is that evolutionarily covarying residues identified from the statistical coupling analysis (SCA), by definition, cannot have a high $\Delta G/k_BT^*$ value; thus SCA cannot detect residues in group I.
The efficacy of C$_{B}$ score in identifying microswitches as well as other hotspots is compared with SCA for the case of rhodopsin in the following section and Figure \ref{SCA}.

\item For AR family, most of the residues with low C$_{B}$ but with high $\Delta G$ score (residues belonging to the group III) are distributed around the ligand binding site and in the cytoplasmic side (Figure S2B). 
The high sequence conservation in the ligand binding sites, identified from the MSA of AR subfamily, is consistent with our general notion that adenosine receptors are specific to the adenosine ligand, which allows the receptor to effectively discriminate other ligand types. 
On the other hand, when MSA is carried out for the entire sequences of the class A GPCR family, the conserved residues are identified more at the cytoplasmic region where G-protein binds (Figure \ref{microswitch}B). 
These findings suggest that the subtype specificity or functional classification is correctly captured in residues with high $\Delta G$ value  as long as a good MSA is used.  

\item There are slight differences in the $C_B$ scores between the apo and agonist-bound forms. 
The contribution of residues satisfying the condition $|C_B^{Ago}-C_B^{Apo}|\geq  0.02$ comes from the group I (12.5 \%, residue number: 197, 246, 278), group II (37.5 \%, residue number: 89, 93, 95, 97, 98, 112, 125, 135, 247), and group IV (50 \%, residue number: 90, 107, 190, 192, 226, 230, 231, 235, 279, 280, 284, 291) (Figure \ref{CB_diff}). 
The residues identified with high $|C_B^{Ago}-C_B^{Apo}|\geq 0.02$ values are mainly located in TM3 and TM5-7. 
Of particular note is that majority of the residues with $|C_B^{Ago}-C_B^{Apo}|\geq 0.02$, also satisfying $C_B^{Apo}\geq 0.05$, are found in the group II (9 out of 12, these residues are marked with underlines in the table of Figure \ref{CB}D), which suggests that among the allosteric hotspots (groups I and II) the less conserved residues (group II) are more sensitive to the $apo\rightarrow ago$ (or inactive$\rightarrow$active) conformational change.  
\end{enumerate}

As presented above, the C$_{B}$-based network analysis of the A$_{2A}$AR structure enables us to identify the allosteric hotspots of A$_{2A}$AR that show neither the sequence variation nor a detectable conformational change in the transition from the apo to agonist-bound form. Next, we will show that the performance of $C_B$-based analysis in identifying the location of microswitches is remarkable by making quantitative comparisons with other conventional approaches.\\

\noindent{\bf Comparison with other approaches.}

\noindent{\it Statistical coupling analysis (SCA)}:
A strong signal of covariation between two remote residues in a multiple sequence alignment, 
which is exploited as a basic principle to identify clusters of residues under long-range coupling in a bioinformatical method called statistical coupling analysis (SCA) \cite{Lockless99Science,Suel2002NSMB,Dima06ProtSci,morcos2011PNAS}, is viewed as a consequence of allosteric communication mediated by multiple groups of residues that lie in the midst of signaling pathways. While it was proposed that the method using SCA on GPCR identified the ``sparse network of coevolving amino acids" (or sectors) \cite{halabi2009cell,McLaughlin2012Nature} that bridges the ligand-binding site to the cytoplasmic G-protein interaction site, forming the allosteric signaling pathways \cite{Suel2002NSMB,Dima06ProtSci}, it fails to detect several highly conserved microswitches. Figure \ref{SCA} shows the list of allosteric hotspots identified for bovine rhodopsin by SCA from two different studies (Figures \ref{SCA}A and \ref{SCA}B) and the residues with high $C_{B} (\geq 0.05)$ (Figure \ref{SCA}C). Although the two methods are based on entirely different assumptions, one solely based on sequence information, the other on network topology, allosteric hotspots identified for rhodopsin are mainly distributed around the TM region. It should, however, be noted that C$_{B}$-based network analysis is much more efficacious in identifying the microswitches, which are considered critical in the activation mechanism of class A GPCRs. 
For rhodopsin, SCA using two slightly different definitions of $\Delta G/k_BT^*$ in Ref.\cite{Suel2002NSMB} and Ref.\cite{Dima06ProtSci} identifies 2 and 5 microswitches, respectively, whereas our $C_B$-based analysis identifies 8 microswitches out of 18  predicted residues. 

Statistical assessment of three results in Figure \ref{SCA} can be made by calculating $\varphi_m$, the ratio between the number of predicted microswitches ($N_m$) and the expectation value ($\langle n\rangle_{rand}$) (Eq.\ref{eqn:varphi}). 
The number of correctly identified microswitches ($N_m$) and the number of residues selected for the prediction ($n$) in each method are $(N_m,n)=(2,31)$, $(5,55)$, and $(8,38)$ for (i) Suel \emph{et al.}, (ii) Dima \emph{et al.}, and (iii) $C_B$-based analysis, respectively. 
Therefore, $\varphi_m^{(i)}=2/1.8$, $\varphi_m^{(ii)}=5/3.2$, and $\varphi_m^{(iii)}=8/2.2$. Note that $(\varphi_m^{(i)}<\varphi_m^{(ii)}<\varphi_m^{(iii)})$ indicates that 
prediction of microswitches made by $C_B$-based analysis is better than those made by SCA; hence attesting to the utility of $C_B$-based analysis.  
\\

\noindent{\it Structural perturbation method (SPM).}
The SPM is used to identify key residues controlling the conformational dynamics by assessing the importance of a residue in the elastic network representation under local perturbation \cite{Zheng05Structure,Zheng06PNAS}. 
The perturbation is invoked by changing the force constant of the springs that link the residue and its neighbors. 
When the overlap of mode $M$ ($\vec{v}_M$) with the vector defining the transition of apo to agonist-bound form ($\vec{r}_{apo\rightarrow ago}=\vec{R}_{ago}-\vec{R}_{apo}$) is significant, i.e., when $\cos{(\vec{r}_{apo\rightarrow ago}\cdot \vec{v}_M)}$ is large, the frequency change of a mode $M$ under the perturbation of $i$-th residue is calculated using $\delta \omega(M,i)=\vec{v}_M^T\cdot \delta H\cdot\vec{v}_M$, where $\delta H$ is the Hessian matrix of the following perturbed energy potential for elastic network model: $\delta E_{ENM}=\frac{1}{2}\sum_{ij}\delta k_o(r_{ij}-r^o_{ij})^2\Theta(r_{ij}^o-R_c$). 
Note that the expression of $\delta\omega(M,i)=\vec{v}_M^T\cdot \delta H\cdot\vec{v}_M$ is analogous to the first-order energy correction term for the $M$-th eigenmode in non-degenerate perturbation theory \cite{Zheng05Structure,Zheng06PNAS}. 
Thus, if a perturbation on the $i$-th residue leads to a large change in $\delta \omega(M,i)$, the residue $i$ is considered to be important in the SPM.

\begin{figure}[!ht]
\begin{center}
\includegraphics[width=5.5in]{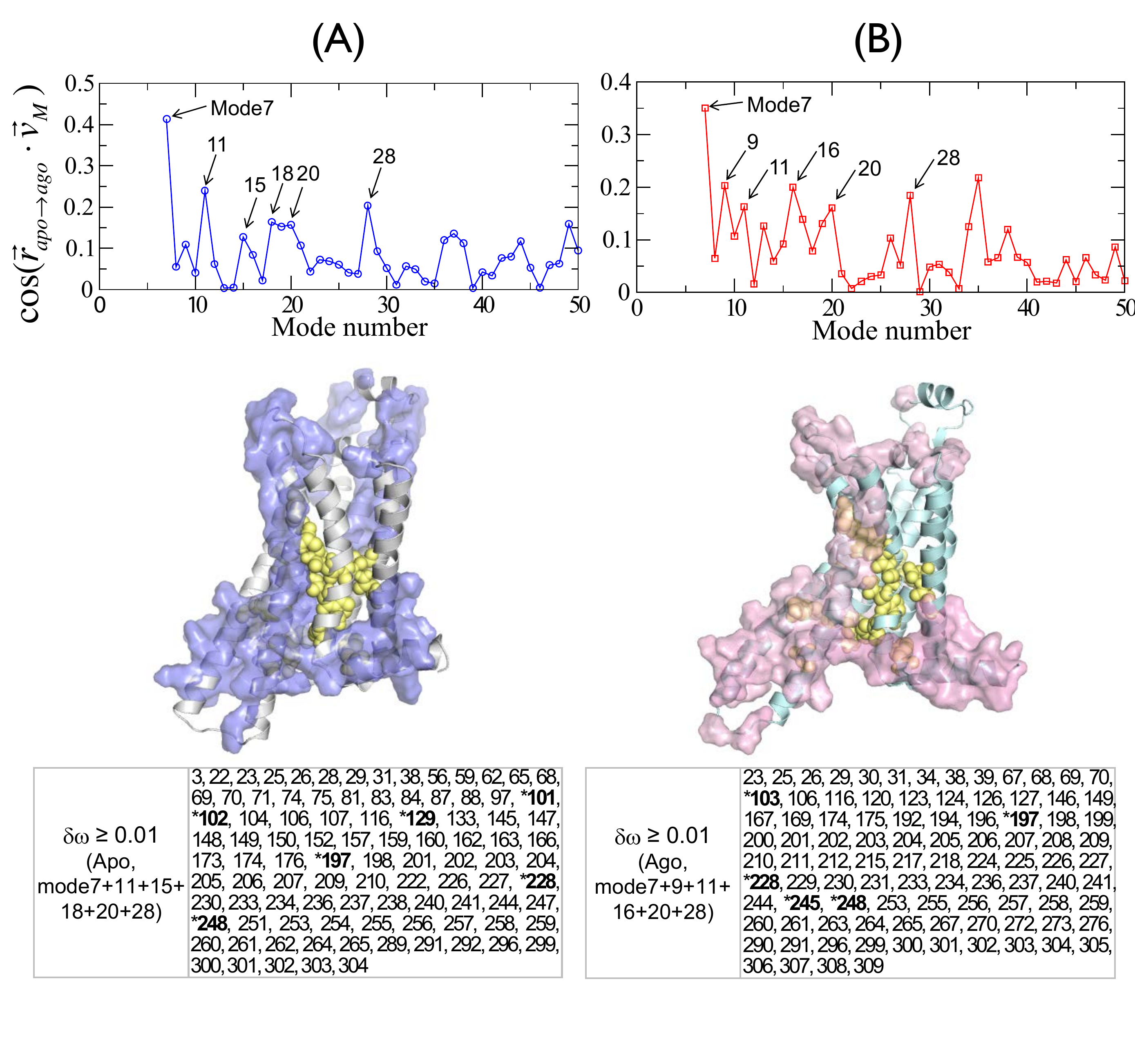}
\end{center}
\caption{
{\bf SPM-identified residues with high $\delta\omega$ values for the superposition of high overlapping modes ($M=7-28$) of (A) the apo (blue) and (B) the agonist-bound form (red).}
The degree of overlap, $\cos{(\vec{r}_{apo\rightarrow ago}\cdot \vec{\nu}_M)}$, calculated between the conformational change from apo to agonist-bound state and M-th normal mode (top).  
 The superposition of hotspot residues, satisfying $\delta \omega(M,i)\geq 0.01$ for high overlapping modes, are depicted with blue and magenta surfaces, respectively, and their residue numbers are listed in the table below, in which the microswitches are marked with asterisks. 
 Note that the key residues identified by SPM are mainly located around the hinge region controlling the motion of TM5-ICL3-TM6. For comparison, the locations of the microswitch residues are depicted with yellow spheres.}
\label{SPM}
\end{figure}

We found that in both for apo and agonist-bound sttructure the mode 7 (excluding the 3 translational and 3 rotational modes, the mode 7 is the lowest eigenmode) has maximum overlap with the conformational change $\vec{r}_{apo\rightarrow ago}$ (Figure \ref{SPM}).
As shown in Figure \ref{SPM}, key residues with high $\delta\omega$ are mainly distributed in the extracellular and intracellular regions of TM helices, which are accompanied with large conformational changes when the transition occurs from the apo to agonist-bound form. Note, however, that even the superposition of six major modes, which have large overlap with conformational changes, is not good enough to identify microswitches that are buried deep inside the GPCR structure.  
Prediction efficiency of SPM that identifies $N_m=6$ and $5$ microswitches out of $n=98$ and $97$ residues for apo and agonist-bound forms (Figure \ref{SPM}) is only   
$\varphi_m=1.1$ and $0.9$, respectively (see Eq.\ref{eqn:varphi}). 
Whereas, $\varphi_m=4.8$ for $C_B$-based analysis indicates that $C_B$-based analysis certainly outperforms SPM in identifying microswitches.   
Hence, neither is the SPM suitable for identifying the microswitches of GPCRs, which undergo only a minor change in their positions before and after the activation. 
\\

\begin{figure}[!ht]
\begin{center}
\includegraphics[width=4.5in]{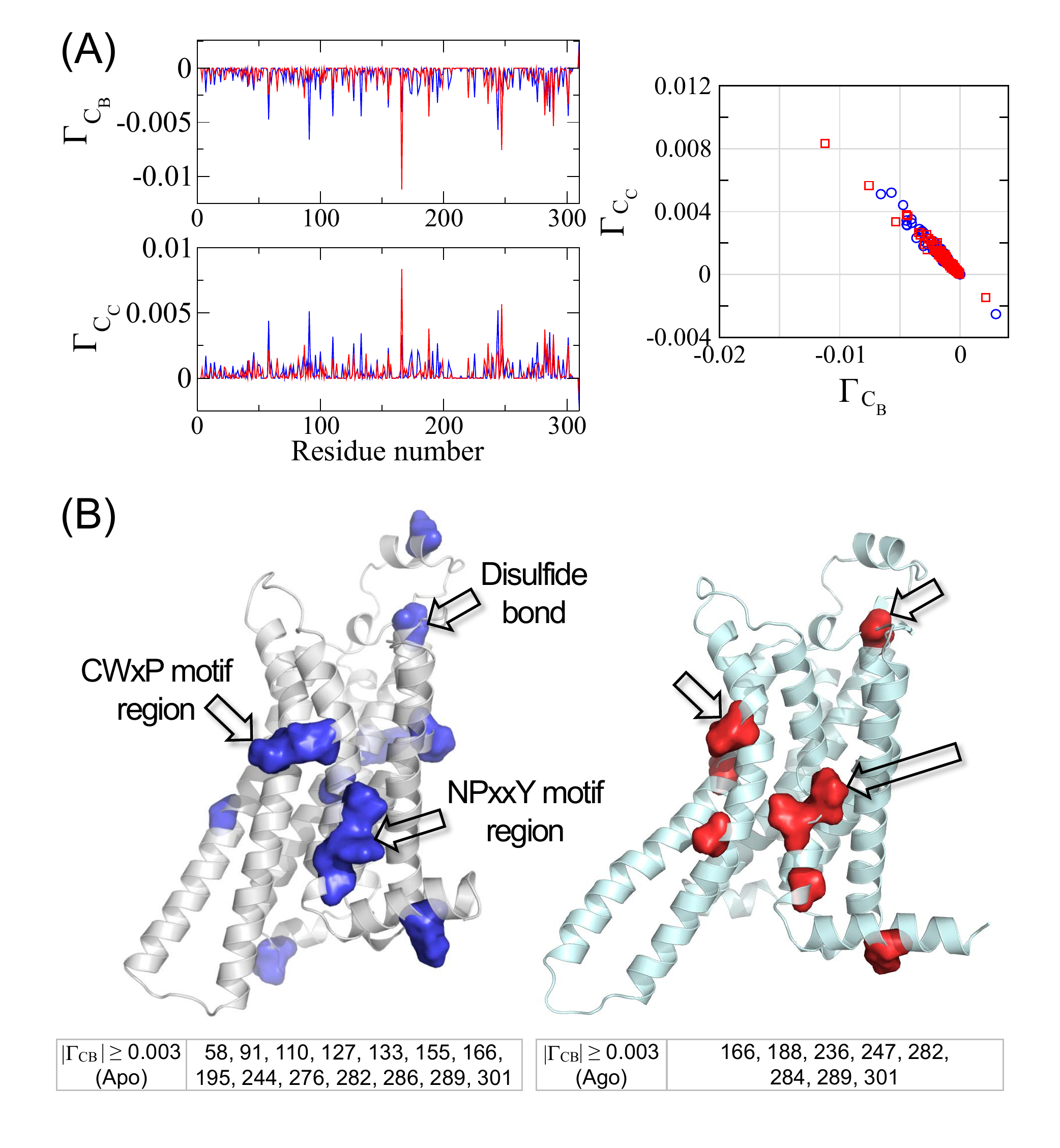}
\end{center}
\caption{
{\bf Glycine scanning (network vulnerability) of A$_{2A}$ adenosine receptor.}  
(A) $\Gamma_{C_B}$ (top) and $\Gamma_{C_C}$ (bottom) (blue: the apo structure; red: the agonist-bound structure). Scatter plot of ($\Gamma_{C_B}$, $\Gamma_{C_C}$) is shown to indicate that $\Gamma_{C_B}$ and $\Gamma_{C_C}$ are well correlated. (B) Regions with high network vulnerability ($|\Gamma|\geq 0.003$) in the apo (left) and agonist-bound forms (right) are represented with blue and red surfaces, and corresponding residue indices are listed in the table.
}
\label{Gly}
\end{figure}

\noindent{\bf The microswitches are critical for the integrity of signaling network of GPCRs.}
In theory of complex networks, a network's tolerance to an error or vulnerability to an attack is evaluated using the relative change in the average network centrality when a node, say $x$, is removed \cite{albert2000Nature}, which can be written as follows: 
\begin{align}
\Gamma_{\xi}^{x}=\frac{\langle C_{\xi}\rangle-\langle C_{\xi}^{x}\rangle}{\langle C_{\xi} \rangle}
\end{align}
where $\langle C_{\xi}\rangle(\equiv\sum_{i=1}^NC_{\xi}(i)/N)$ is the average network centrality, and $\langle C_{\xi}^{x}\rangle$  is a value evaluated for a newly constructed network when the node $x$ is removed from the original network. 
The idea of network vulnerability is, in fact, routinely practiced in molecular biology in the form of protein mutagenesis assay, which measures the effect of mutations on the degree to which proteins can retain their activity. 
Adapting the idea of network vulnerability, we performed in silico glycine scanning of the constructed residue interaction network of the A$_{2A}$AR. As straightforwardly implicated by the term ``glycine scanning", we mimicked the protein mutagenesis assay by deleting the side chain of each residue and evaluated the deletion effect on the network. 
Our glycine scanning analysis differs from the previous study applying network analysis \cite{delSol2006MSB} in that only a side chain, rather than the entire residue, is deleted for each scan. It is important to keep C$\alpha$ backbone because, even in the absence of the side chain, intra-molecular residue contacts can still be formed via backbone-side chain or backbone-backbone interactions. 
Note that here a readjustment of local environment due to the side chain removal is not considered. 
Our aim here is to make a quantitative assessment of the role of the side chain in the original residue interaction network. The greater is the role played by the removed side chain in maintaining the network structure, the more significant would be the response of average network centrality to the removal of that particular residue.
We assess the effect of deleting side chains by calculating the changes in average closeness ($\langle C_C\rangle$) or betweenness centralities ($\langle C_B\rangle$), both of which turn out to be highly correlated (Figure \ref{Gly}A).

Our glycine scanning analysis identified the group of residues critical for the integrity of interaction network that is responsible for the receptor allostery. The residues with strong network vulnerability ($|\Gamma_{C_{\xi}}|\geq 0.003$) are identified in the regions around CWxP and NPxxY motifs (Figure \ref{Gly}B) \cite{jaakola2008Science}, which retain proline that creates a kinked helix in the middle of TM6 or TM7 \cite{nygaard2009TPS}. In the inactive state of GPCRs, interactions between the cytoplasmic ends of TM3 and TM6 constrain the relative motion of these segments by forming an ionic-lock between R102$^{3.50}$ and E228$^{6.30}$ \cite{nygaard2009TPS}. Disruption of such constraint, triggered by agonist binding, enables TM6 to move outward from TM3 (see DRY motif \& ionic lock in the Figure \ref{microswitch}C). NPxxY motif, which interacts with TM6 or helix 8, imposes structural constraints in GPCRs and stabilizes the helical structures \cite{shi2002JBC,fritze2003PNAS}.  In addition, C166, which constrains ECL1 and ECL2 by forming a disulfide bond with C77, is detected to have high network vulnerability. It is of note that the constrained random coil structure of ECL2 is unique to A$_{2A}$AR in that the ECL2 of other GPCRs typically forms $\beta$ sheet or $\alpha$-helix \cite{jaakola2008Science}.\\

\noindent{\bf Distinct $C_B$-based wiring diagrams reflect GPCR subtype specificity. }
Here we extend the $C_B$-based network analysis to other class A GPCRs, including $\beta_1$, $\beta_2$ adrenergic receptors (PDB IDs: 2VT4 and 3NYA), chemokine CXCR4 receptor (3ODU), dopamine D3 receptor (3PBL), histamine H1 receptor (3RZE), and bovine rhodopsin (1U19) \cite{katritch2011TPS}. Similar to the A$_{2A}$AR, the network of residues with high $C_{B} (\geq 0.05)$ in these class A GPCRs form contiguous surface that bridges between the ligand binding and G-protein binding sites (Figure S4). In most GPCRs the high C$_{B}$ residues are mainly distributed around the ``minor binding pocket", located in the shallow part of the ligand binding site between the TM1, 2, 3 and TM7, which serves as an onset point of orthosteric signal transduction process \cite{Rosenkilde2010TPS}. In particular, when the $C_B$ is restricted to a value greater than 0.075, the high C$_{B}$ residues bridge the extracellular region of TM3 to the TM6, 7 helices. For the class A GPCRs, the highly vulnerable residues identified by the glycine scanning analysis are mostly distributed in TM3 and TM7 (Figure S5). Note that K296$^{7.43}$ in bovine rhodopsin, known to contribute to the activation of rhodopsin by forming a covalent bond with retinal \cite{Rosenkilde2010TPS}, is also identified (the residue marked with a yellow arrow in Figure S5G). Along with the variation of residues (F168 in A$_{2A}$AR; R183 and Y190 in CXCR4 receptor; and K179 in H1 receptor) and wiring diagram in ECL2 detected by glycine scanning analysis, the variations in the high C$_{B}$-surfaces demonstrated in the class A GPCRs (Figure S4) are deemed responsible for their subtype selectivity \cite{Rosenkilde2010TPS}. \\

\begin{figure}[!ht]
\begin{center}
\includegraphics[width=6in]{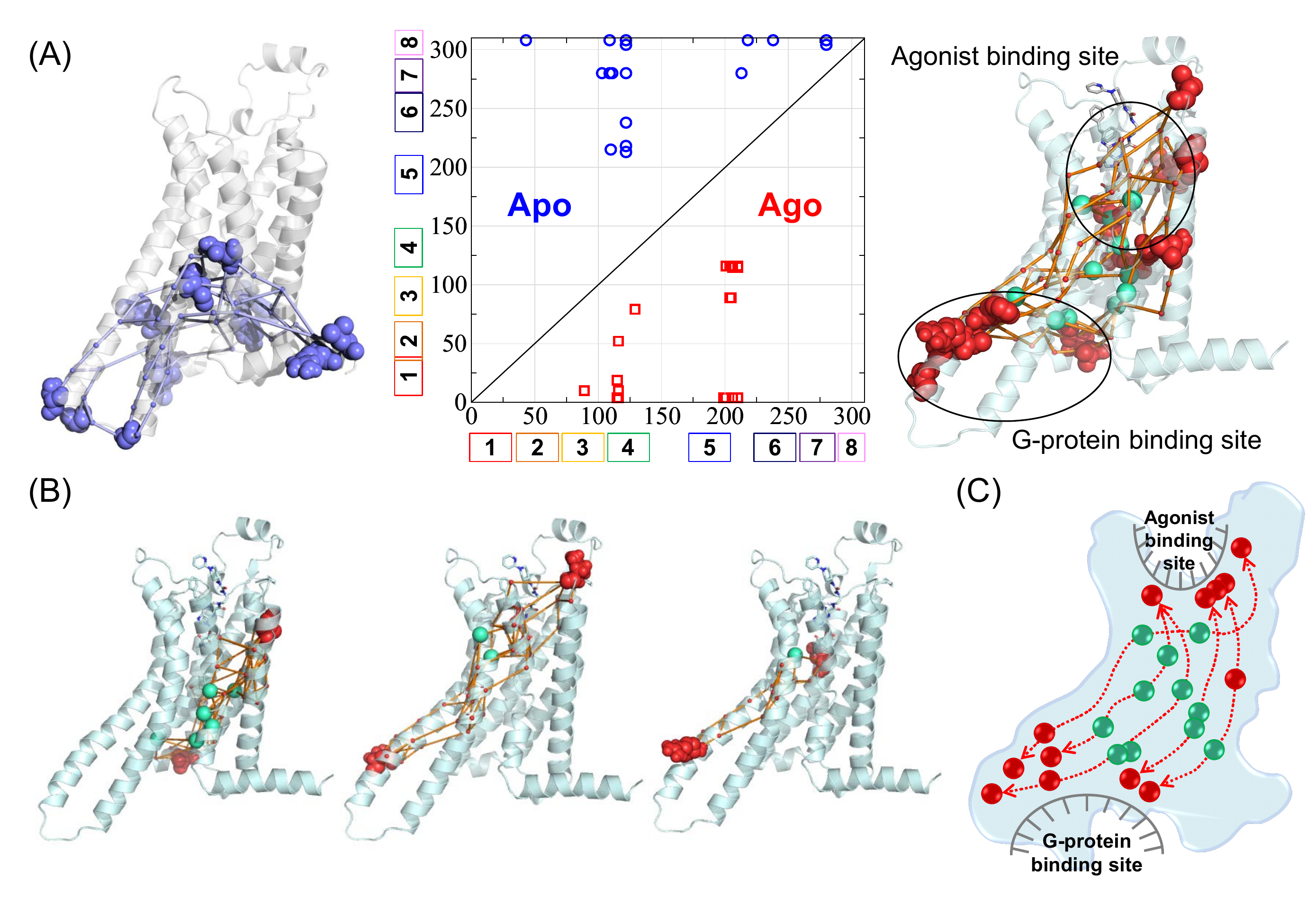}
\end{center}
\caption{
{\bf Multiple pathways of intramolecular signals that link the cross-correlated residues in the extra and intra-cellular domains of A$_{2A}$AR.}  
(A) Residue pairs with high cross-correlation ($|CC_{ij}| \geq 0.5$) and distance greater than 6 ($d_{ij} > 6$) are marked using blue circles and red squares for the apo form (upper left corner) and for the agonist-bound form (lower right corner), respectively (see the original map in Figure S6). The minimum paths between the cross-correlated residues are shown for apo and agonist-bound forms on the left and right, respectively. For agonist-bound form, long-range cross-correlations are detected between the extracellular ligand binding and cytoplasmic G protein binding sites. Microswitches and other residues with high $C_{B}$ on the paths are displayed in cyan spheres. (B) Examples of the multiple signaling paths between the agonist binding and G-protein binding sites. (C) Schematic of transmembrane signaling represented by multiple shortest paths linking the long-range cross-correlated residues. }
\label{crosscorr}
\end{figure}

\noindent{\bf Long-range transmembrane cross-correlation in the agonist-bound active state.}
As suggested by the G-protein bound structure \cite{Rosenbaum2009Nature}, it is expected that the agonist binding site in the extracellular side and intracellular region are functionally coupled in the active forms of GPCRs, and this coupling is mediated by a structural reorganization of seven membered TM helices. 
To quantify such long-range coupling in the dynamics of A$_{2A}$AR, we calculated cross-correlation between residues in terms of $C_B$ (Eq. 2) by using the conformational ensemble of the A$_{2A}$AR generated from the 300 nsec MD simulation trajectory (see Materials and Methods and Figure S9). 
\begin{align}
CC_{ij}=\frac{\langle C_B(i)C_B(j)\rangle-\langle C_B(i)\rangle\langle C_B(j)\rangle}{\sqrt{\langle(\delta C_B(i))^2\rangle}\sqrt{\langle(\delta C_B(j))^2\rangle}}
\label{eqn:CC}
\end{align}
where $\langle\ldots\rangle$ refers to an ensemble average; thus $\langle C_B(i)\rangle$ denotes the average betweenness centrality for the residue $i$. As shown in the cross-correlation matrices for apo and agonist-bound form (Figures S6 and S7), the signatures of correlation between residues are scattered all over the structure.  
To identify residue pairs with long-range cross-correlation we imposed the conditions of $CC_{ij} \geq 0.5$ and $d_{ij}>6$ (Figure \ref{crosscorr}A).
Importantly, while in the apo structure the residue pairs under high cross-correlations are distributed only around the cytoplasmic side of the TMs (Figures \ref{crosscorr}A, S7), functionally important long-range couplings are detected between the ligand-binding and cytoplasmic G-protein binding sites in the agonist-bound form (Figures \ref{crosscorr}A, S7). 
This result from the agonist-bound form is consistent with the view that a bound agonist makes tight interactions with the surrounding residues and increases the receptor activity above its basal level \cite{Rosenbaum2009Nature}.  
The long-range coupling between the ligand binding site and G-protein binding site for the agonist-bound form is also grasped by computing the mean square fluctuation using structural ensemble (see Figure S8).

Notably, there are multiple parallel paths linking the correlated residues \cite{delSol09Structure}, the degeneracy of which varies from 1 to as many as 480 depending on the residue pair 
(For the details of entire paths between the correlated residues, see Part 1 and 2 in Supporting Information II). 
The presence of multiple parallel pathways is consistent with the recent new view of allostery \cite{daily2009PLoSCB,cui2008ProtSci,delSol09Structure}. 
As some of the representative allosteric paths, linking the residues in extracellular and intracellular regions, are demonstrated in Figure \ref{crosscorr}B, 
the 80 \% of transmembrane signaling paths go through the residues with high C$_{B}$, which includes the microswitches as well as other functionally important residues (see the residues represented with cyan spheres in Figures \ref{crosscorr}B, \ref{crosscorr}C and Part 1, 2 in Supporting Information II). 
It is these residues, lying in the midst of communication pathways, that toggle the intra-molecular signaling. 
The qualitatively disparate results displayed in apo and agonist-bound forms provide a picture consistent with the function of GPCRs. 

To systematically group correlated residues, we carried out hierarchical clustering analysis on the acquired matrices and represented the result using dendrogram (Figure S7). From the two clusters of positively correlated residues (cluster 1 and 2), the clusters of residue pairs with the strongest signal are shown on each clustered cross-correlation map of the apo and agonist-bound form with residue indices. 
The cross-correlated residue paris obtained using hierarchical clustering analysis (Figure S7) are similar to those from the simple condition of $CC_{ij}\geq 0.5$ and $d_{ij}>6$ that we imposed in Figure \ref{crosscorr}. 
The residues within each of cluster 1 and 2 are the parts of structure that ``breathe together" in terms of $C_B$ values. Also, it is noteworthy that in terms of the correlation of $C_B$ value there is a strong anti-correlation between cluster 1 and cluster 2, which suggests that "breathing" of residues in cluster 1 and cluster 2 occurs out-of-phase.

Lastly, it is worth considering the signaling paths on a weighted graph. To this end, we employed the ``dynamical network community analysis" \cite{Sethi09PNAS}, implemented to the molecular visualization package VMD. 
In this analysis, an inter-residue cross-correlation calculated from an ensemble of structures from long MD simulation is used for the weight of edges in the network.    
Using the NetworkView module in VMD and our 300 ns MD simulation as an  input, we calculated the optimal and a set of suboptimal paths (offset$=20$) between residue pairs that show long-range cross-correlation (Supporting Information II - Part 3).  
In most of the cases, the allosteric signaling paths computed on unweighted and weighted graphs for agonist-bound form are qualitatively similar; yet, it is interesting to point out the large detour in the signaling paths of the residue pairs 116-4 and 116-10 on weighted graph (see Figure S5 in Supporting Information II).

\section*{Conclusions}
Deciphering the protein allostery has long been one of the grand challenges in molecular, structural, and computational biology. We elucidated allosteric hotspots and signaling pathways of the A$_{2A}$AR and other class A GPCRs by using the measure of betweenness centrality for each residue in protein structure network, the glycine scanning analysis, and the cross-correlation analysis based on the structural ensemble from MD simulations. 
Just like the role of native topology has been illuminated in the folding and unfolding mechanisms of proteins \cite{OnuchicARPC97,Alm99PNAS,Klimov00PNAS,Ferrara01JMB}, the success of analysis using graph representations of protein topology underscores the importance of native protein topology as one of the most critical determinants for intramolecular allosteric signaling. It is of special note that signals generated from protein dynamics, which include changes of inter-residue force, contact, or even local packing, are transmitted via the contacts formed between two neighboring residues. 
From the perspective of signal transduction, the betweenness centrality, defined with the number of parallel pathways on a given node,  
is physically a sensible way to quantify the amount of traffic on the node, thus to identify allosteric hotspots for a given protein structure. 
Given that residues of GPCRs associated with allostery and their signaling pathways are hard to capture using other conventional methods exploiting the information of sequence coevolution or variants of normal mode analysis (Figures \ref{SCA} and \ref{SPM}), the success of $C_{B}$-based analysis presented here is remarkable.

At the current stage, not only in the context of allosteric modulations in drug design \cite{Swain2006COSB,goodey2008NCB,hardy2004COSB} but also in the ligand binding (or release) induced conformational change in biological motors \cite{Hyeon06PNAS,Hyeon11BJ}, the importance of allostery in understanding the protein dynamics is highlighted more than ever. 
From the methodological perspective of this study, our $C_B$-based network analysis on protein structures is found quite powerful in identifying allosteric hotspots, and the results of analysis are in strong correlation with biochemical studies. The list of key residues for allostery and their cross-correlation identified here should be of great help to design experiments as well as contribute to our understanding to the dynamics of GPCRs.
Our simple approach can not only be extended to study the allostery of other important proteins but also to study the allosteric communication within protein-protein or protein-RNA complexes \cite{Sethi09PNAS}. 

\section*{Acknowledgments}
This work was supported by the grant from the National Leading Research Lab (NLRL) program (2011-0028885) funded by the Ministry of Education, Science and Technology (MEST) and the National Research Foundation of Korea (NRF). 
We thank Korea Institute for Advanced Study (KIAS) and Korea Institute of Science and Technology Information (KISTI) Supercomputing Center for providing computing resources.


\begin{thebibliography}{10}

\bibitem{Rosenbaum2009Nature}
Rosenbaum, D., Rasmussen, S., and Kobilka, B.
\newblock {The structure and function of G-protein-coupled receptors}.
\newblock Nature,  2009;459(7245):356--363.

\bibitem{Jensen2004EJPS}
Jensen, A. and Spalding, T.
\newblock {Allosteric modulation of G-protein coupled receptors}.
\newblock Eur. J. Pharma. Sci. 2004;21(4):407--420.

\bibitem{wess1997FASEBJ}
Wess, J.
\newblock {G-protein-coupled receptors: molecular mechanisms involved in
  receptor activation and selectivity of G-protein recognition}.
\newblock FASEB J. 1997;11(5):346--354.

\bibitem{Kjelsberg1992JBC}
Kjelsberg, M., Cotecchia, S., Ostrowski, J., Caron, M., and Lefkowitz, R.
\newblock {Constitutive activation of the alpha 1B-adrenergic receptor by all
  amino acid substitutions at a single site. Evidence for a region which
  constrains receptor activation}.
\newblock J. Biol. Chem. 1992;267(3):1430--1433.

\bibitem{Bond06TPS}
Bond, R.~A. and Ijzerman, A.~P.
\newblock {Recent developments in constitutive receptor activity and inverse
  agonism, and their potential for GPCR drug discovery}.
\newblock Trends Pharmacol. Sci. 2006;27:92--96.

\bibitem{Yun08PNAS}
Yun, C.~H., Mengwasser, K.~E., Toms, A.~V., Woo, M.~S., Greulich, H., Wong,
  K.~K., Meyerson, M., and Eck, M.~J.
\newblock {The T790M mutation in EGFR kinase causes drug resistance by
  increasing the affinity for ATP}.
\newblock Proc. Natl. Acad. Sci. U.S.A. 2008;105:2070--2075.

\bibitem{Yonath05ARBiochem}
Yonath, A.
\newblock Antibiotics targeting ribosomes: resistance, selectivity, synergism,
  and cellular regulation.
\newblock Annu. Rev. Biochem. 2005;74:649--679.

\bibitem{Galzi94COSB}
Galzi, J.~L. and Changeux, J.~P.
\newblock Neurotransmitter-gated ion channels as unconventional allosteric
  proteins.
\newblock Curr. Opin. Struct. Biol. 1994;4:554--565.

\bibitem{Changeux2012ARB}
Changeux, J.~P.
\newblock Allostery and the Monod-Wyman-Changeux Model After 50 Years.
\newblock Annu. Rev. Biophys. 2012;41:103--133.

\bibitem{liu2012Science}
Liu, W., Chun, E., Thompson, A., Chubukov, P., Xu, F., Katritch, V., Han, G.,
  Roth, C., Heitman, L., IJzerman, A., Cherezov, V., and Stevens, R.
\newblock {Structural basis for allosteric regulation of GPCRs by sodium ions}.
\newblock Science 2012;337:232--236.

\bibitem{Lockless99Science}
Lockless, S.~W. and Ranganathan, R.
\newblock Evolutionarily Conserved Pathways of Energetic Connectivity in
  Protein Families.
\newblock Science 1999;286:295--299.

\bibitem{Suel2002NSMB}
S{\"u}el, G., Lockless, S., Wall, M., and Ranganathan, R.
\newblock Evolutionarily conserved networks of residues mediate allosteric
  communication in proteins.
\newblock Nature Struct. Mol. Biol. 2002;10(1):59--69.

\bibitem{Dima06ProtSci}
Dima, R.~I. and Thirumalai, D.
\newblock Determination of network of residues that regulate allostery in
  protein families using sequence analysis.
\newblock Protein Sci. 2006;15:258--268.

\bibitem{morcos2011PNAS}
Morcos, F., Pagnani, A., Lunt, B., Bertolino, A., Marks, D., Sander, C.,
  Zecchina, R., Onuchic, J., Hwa, T., and Weigt, M.
\newblock Direct-coupling analysis of residue coevolution captures native
  contacts across many protein families.
\newblock Proc. Natl. Acad. Sci. U.S.A. 2011;108(49):E1293--E1301.

\bibitem{Zheng05Structure}
Zheng, W., Brooks, B.~R., Doniach, S., and Thirumalai, D.
\newblock Network of Dynamically Important Residues in the Open/Closed
  Transition in Polymerases Is Strongly Conserved.
\newblock Structure 2005;13:565--577.

\bibitem{Zheng06PNAS}
Zheng, W., Brooks, B.~R., and Thirumalai, D.
\newblock Low-frequency normal modes that describe allosteric transitions in
  biological nanomachines are robust to sequence variations.
\newblock Proc. Natl. Acad. Sci. U.S.A. 2006;103:7664--7669.

\bibitem{BaharMSB06}
Chennubhotla, C. and Bahar, I.
\newblock {Markov propagation of allosteric effects in biomolecular systems:
  application to GroEL-GroES}.
\newblock Mol. Syst. Biol. 2006;2:msb4100075.

\bibitem{balabin2009PNAS}
Balabin, I., Yang, W., and Beratan, D.
\newblock Coarse-grained modeling of allosteric regulation in protein
  receptors.
\newblock Proc. Natl. Acad. Sci. U.S.A. 2009;106(34):14253--14258.

\bibitem{Lee11JACS}
Lee, Y., Jeong, L.~S., Choi, S., and Hyeon, C.
\newblock {Link between Allosteric Signal Transduction and Functional Dynamics
  in a Multisubunit Enzyme: S-Adenosylhomocysteine Hydrolase}.
\newblock J. Am. Chem. Soc. 2011;133:19807--19815.

\bibitem{nygaard2009TPS}
Nygaard, R., Frimurer, T., Holst, B., Rosenkilde, M., and Schwartz, T.
\newblock Ligand binding and micro-switches in {7TM} receptor structures.
\newblock Trends Pharmacol. Sci. 2009;30(5):249--259.

\bibitem{freeman1979SocialNetworks}
Freeman, L.
\newblock Centrality in social networks conceptual clarification.
\newblock Social networks 1979;1(3):215--239.

\bibitem{newman2005SocialNetworks}
Newman, M.
\newblock A measure of betweenness centrality based on random walks.
\newblock Social networks 2005;27(1):39--54.

\bibitem{Jeong01Nature}
Jeong, H., Mason, S.~P., Barabasi, A.-L., and Oltvai, Z.~N.
\newblock Lethality and centrality in protein networks.
\newblock Nature 2001;411:41--42.

\bibitem{delSol2006MSB}
Del~Sol, A., Fujihashi, H., Amoros, D., and Nussinov, R.
\newblock Residues crucial for maintaining short paths in network communication
  mediate signaling in proteins.
\newblock Mol. Sys. Biol. 2006;2:2006.0019.

\bibitem{edgar2004muscle}
Edgar, R.
\newblock MUSCLE: a multiple sequence alignment method with reduced time and
  space complexity.
\newblock BMC bioinformatics 2004;5(1):113.

\bibitem{xu2011Science}
Xu, F., Wu, H., Katritch, V., Han, G., Jacobson, K., Gao, Z., Cherezov, V., and
  Stevens, R.
\newblock {Structure of an agonist-bound human A2A adenosine receptor}.
\newblock Science 2011;332(6027):322.

\bibitem{jaakola2008Science}
Jaakola, V., Griffith, M., Hanson, M., Cherezov, V., Chien, E., Lane, J.,
  IJzerman, A., and Stevens, R.
\newblock {The 2.6 angstrom crystal structure of a human A2A adenosine receptor
  bound to an antagonist}.
\newblock Science 2008;322:1211--1217.

\bibitem{deupi2011COSB}
Deupi, X. and Standfuss, J.
\newblock Structural insights into agonist-induced activation of
  G-protein-coupled receptors.
\newblock Curr. Opin. Struct. Biol. 2011;21:541--551.

\bibitem{lebon2011Nature}
Lebon, G., Warne, T., Edwards, P., Bennett, K., Langmead, C., Leslie, A., and
  Tate, C.
\newblock {Agonist-bound adenosine A2A receptor structures reveal common
  features of GPCR activation}.
\newblock Nature 2011;474(7352):521--525.

\bibitem{Buck2006BiophysJ}
Buck, M., Bouguet-Bonnet, S., Pastor, R.~W., and MacKerell~Jr, A.~D.
\newblock Importance of the CMAP correction to the CHARMM22 protein force
  field: dynamics of hen lysozyme.
\newblock Biophys. J. 2006;90(4):L36--L38.

\bibitem{da2009protein}
da~Silveira, C., Pires, D., Minardi, R., Ribeiro, C., Veloso, C., Lopes, J.,
  Meira~Jr, W., Neshich, G., Ramos, C., Habesch, R., and Marcelo, M.
\newblock Protein cutoff scanning: A comparative analysis of cutoff dependent
  and cutoff free methods for prospecting contacts in proteins.
\newblock Proteins: Structure, Function, and Bioinformatics
  2009;74(3):727--743.

\bibitem{Watts98Nature}
Watts, D.~J. and Strogatz, S.~H.
\newblock Collective dynamics of 'small-world' netoworks.
\newblock Nature 1998;393:440--442.

\bibitem{albert2000Nature}
Albert, R., Jeong, H., and Barab{\'a}si, A.
\newblock Error and attack tolerance of complex networks.
\newblock Nature 2000;406(6794):378--382.

\bibitem{jeong2000Nature}
Jeong, H., Tombor, B., Albert, R., Oltvai, Z., and Barab{\'a}si, A.
\newblock The large-scale organization of metabolic networks.
\newblock Nature 2000;407(6804):651--654.

\bibitem{balazsi2005PNAS}
Balazsi, G., Barab{\'a}si, A., and Oltvai, Z.
\newblock {Topological units of environmental signal processing in the
  transcriptional regulatory network of Escherichia coli}.
\newblock Proc. Natl. Acad. Sci. U.S.A. 2005;102:7841--7846.

\bibitem{vendruscolo2002PRE}
Vendruscolo, M., Dokholyan, N., Paci, E., and Karplus, M.
\newblock Small-world view of the amino acids that play a key role in protein
  folding.
\newblock Phys. Rev. E. 2002;65(6):061910.

\bibitem{atilgan2004BJ}
Atilgan, A., Akan, P., and Baysal, C.
\newblock Small-world communication of residues and significance for protein
  dynamics.
\newblock Biophys. J. 2004;86(1):85--91.

\bibitem{greene2003JMB}
Greene, L. and Higman, V.
\newblock Uncovering network systems within protein structures.
\newblock J. Mol. Biol. 2003;334(4):781--791.

\bibitem{amitai2004JMB}
Amitai, G., Shemesh, A., Sitbon, E., Shklar, M., Netanely, D., Venger, I., and
  Pietrokovski, S.
\newblock Network analysis of protein structures identifies functional
  residues.
\newblock J. Mol. Biol. 2004;344(4):1135--1146.

\bibitem{bagler2007Bioinfomatics}
Bagler, G. and Sinha, S.
\newblock Assortative mixing in Protein Contact Networks and protein folding
  kinetics.
\newblock Bioinformatics 2007;23(14):1760--1767.

\bibitem{brinda2005BJ}
Brinda, K. and Vishveshwara, S.
\newblock A network representation of protein structures: implications for
  protein stability.
\newblock Biophys. J. 2005;89(6):4159--4170.

\bibitem{dokholyan2002PNAS}
Dokholyan, N., Li, L., Ding, F., and Shakhnovich, E.
\newblock Topological determinants of protein folding.
\newblock Proc. Natl. Acad. Sci. U.S.A. 2002;99(13):8637.

\bibitem{Sethi09PNAS}
Sethi, A., Eargle, J., Black, A.~A., and Luthey-Schulten, Z.
\newblock Dynamical networks in tRNA:protein complexes.
\newblock Proc. Natl. Acad. Sci. U.S.A. 2009;106:6620--6625.

\bibitem{yao2013BJ}
Yao, X.-Q. and Grant, B.~J.
\newblock Domain-Opening and Dynamic Coupling in the< i> $\alpha$</i>-Subunit
  of Heterotrimeric G Proteins.
\newblock Biophys. J. 2013;105(2):L08--L10.

\bibitem{miao2013PNAS}
Miao, Y., Nichols, S.~E., Gasper, P.~M., Metzger, V.~T., and McCammon, J.~A.
\newblock Activation and dynamic network of the M2 muscarinic receptor.
\newblock Proc. Natl. Acad. Sci. U.S.A. 2013;110:10982?--10987.

\bibitem{Dijkstra59NM}
Dijkstra, E.~W.
\newblock A Note on Two Problems in Connection with Graphs.
\newblock Numerische Math 1959;1:269--271.

\bibitem{brandes2001JMS}
Brandes, U.
\newblock A faster algorithm for betweenness centrality.
\newblock J. Math. Soc. 2001;25(2):163--177.

\bibitem{Atilgan2007BJ}
Atilgan, A.~R., Turgut, D., and Atilgan, C.
\newblock Screened nonbonded interactions in native proteins manipulate optimal
  paths for robust residue communication.
\newblock Biophys. J. 2007;92(9):3052--3062.

\bibitem{park2011BMCbioinfo}
Park, K. and Kim, D.
\newblock Modeling allosteric signal propagation using protein structure
  networks.
\newblock BMC bioinformatics 2011;12(Suppl 1):S23.

\bibitem{daily2009PLoSCB}
Daily, M.~D. and Gray, J.~J.
\newblock Allosteric communication occurs via networks of tertiary and
  quaternary motions in proteins.
\newblock PLoS computational biology 2009;5(2):e1000293.

\bibitem{rasmussen2011Nature}
Rasmussen, S., DeVree, B., Zou, Y., Kruse, A., Chung, K., Kobilka, T., Thian,
  F., Chae, P., Pardon, E., Calinski, D., Mathiesen, J., Shah, S., Lyons, J.,
  Caffrey, M., Gellman, S., Steyaert, J., Skinoitis, G., Weis, W., Sunahara,
  R., and Kobilka, B.
\newblock Crystal structure of the $\beta_2$ adrenergic receptor-Gs protein
  complex.
\newblock Nature 2011;477(7366):549--555.

\bibitem{katritch2013ARPT}
Katritch, V., Cherezov, V., and Stevens, R.~C.
\newblock Structure-function of the G protein-coupled receptor superfamily.
\newblock Annu. Rev. Pharmacol. Toxicol. 2013;53:531--556.

\bibitem{hofmann2009Hofmann}
Hofmann, K., Scheerer, P., Hildebrand, P., Choe, H., Park, J., Heck, M., and
  Ernst, O.
\newblock AG protein-coupled receptor at work: the rhodopsin model.
\newblock Trends Biochem. Sci. 2009;34(11):540--552.

\bibitem{suwa2011Pharmaceuticals}
Suwa, M., Sugihara, M., and Ono, Y.
\newblock {Functional and structural overview of G-protein-coupled receptors
  comprehensively obtained from genome sequences}.
\newblock Pharmaceuticals 2011;4(4):652--664.

\bibitem{ballesteros1995MethodsNeurosci}
Ballesteros, J. and Weinstein, H.
\newblock {Integrated methods for modeling G-protein coupled receptors}.
\newblock Methods Neurosci 1995;25:366--428.

\bibitem{trzaskowski2012CMC}
Trzaskowski, B., Latek, D., Yuan, S., Ghoshdastider, U., Debinski, A., and
  Filipek, S.
\newblock {Action of Molecular Switches in GPCRs-Theoretical and Experimental
  Studies}.
\newblock Curr. Med. Chem. 2012;19(8):1090--1109.

\bibitem{jacobson2005ChemBiol}
Jacobson, K.~A., Ohno, M., Duong, H.~T., Kim, S.~K., Tchilibon, S., Cesnek, M.,
  Hol{\`y}, A., and Gao, Z.~G.
\newblock {A Neoceptor Approach to Unraveling Microscopic Interactions between
  the Human A$_{2A}$ Adenosine Receptor and Its Agonists}.
\newblock Chemistry \& Biology 2005;12(2):237--247.

\bibitem{kim2003JMC}
Kim, S., Gao, Z., Van~Rompaey, P., Gross, A., Chen, A., Van~Calenbergh, S., and
  Jacobson, K.
\newblock {Modeling the adenosine receptors: comparison of the binding domains
  of A2A agonists and antagonists}.
\newblock J. Med. Chem. 2003;46(23):4847--4859.

\bibitem{Horn2003NAR}
Horn, F., Bettler, E., Oliveira, L., Campagne, F., Cohen, F., and Vriend, G.
\newblock {GPCRDB information system for G protein-coupled receptors}.
\newblock Nucleic Acids Res. 2003;31(1):294--297.

\bibitem{angel2009PNAS}
Angel, T.~E., Gupta, S., Jastrzebska, B., Palczewski, K., and Chance, M.~R.
\newblock Structural waters define a functional channel mediating activation of
  the GPCR, rhodopsin.
\newblock Proc. Natl. Acad. Sci. U.S.A. 2009;106(34):14367--14372.

\bibitem{halabi2009cell}
Halabi, N., Rivoire, O., Leibler, S., and Ranganathan, R.
\newblock Protein sectors: evolutionary units of three-dimensional structure.
\newblock Cell 2009;138(4):774--786.

\bibitem{McLaughlin2012Nature}
McLaughlin, R.~N., Poelwijk, F.~J., Raman, A., Gosal, W.~S., and Ranganathan,
  R.
\newblock The Spatial Architecture of Protein Function and Adaptation.
\newblock Nature 2012;138:138--142.

\bibitem{shi2002JBC}
Shi, L., Liapakis, G., Xu, R., Guarnieri, F., Ballesteros, J., and Javitch, J.
\newblock $\beta$2 Adrenergic Receptor Activation.
\newblock J. Biol. Chem. 2002;277(43):40989--40996.

\bibitem{fritze2003PNAS}
Fritze, O., Filipek, S., Kuksa, V., Palczewski, K., Hofmann, K., and Ernst, O.
\newblock {Role of the conserved NPxxY (x) 5, 6F motif in the rhodopsin ground
  state and during activation}.
\newblock Proc. Natl. Acad. Sci. U.S.A. 2003;100(5):2290.

\bibitem{katritch2011TPS}
Katritch, V., Cherezov, V., and Stevens, R.
\newblock Diversity and modularity of G protein-coupled receptor structures.
\newblock Trends Pharmacol. Sci. 2011;33:17--27.

\bibitem{Rosenkilde2010TPS}
Rosenkilde, M.~M., Benned-Jensen, T., Frimurer, T.~M., and Schwartz, T.~W.
\newblock {The minor binding pocket: a major player in 7TM receptor
  activation}.
\newblock Trends Pharmacol. Sci. 2010;31:567--574.

\bibitem{delSol09Structure}
{del Sol}, A., Tsai, C.-J., Ma, B., and Nussinov, R.
\newblock The Origin of Allosteric Functional Modulation: Multiple Pre-existing
  Pathways.
\newblock Structure 2009;17:1042--1050.

\bibitem{cui2008ProtSci}
Cui, Q. and Karplus, M.
\newblock Allostery and cooperativity revisited.
\newblock Protein Science 2008;17(8):1295--1307.

\bibitem{OnuchicARPC97}
Onuchic, J.~N., Luthey-Schulten, Z., and Wolynes, P.~G.
\newblock {Theory of Protein Folding: The Energy Landscape Perspective}.
\newblock Ann. Rev. Phys. Chem. 1997;48:539--600.

\bibitem{Alm99PNAS}
Alm, E. and Baker, D.
\newblock {Prediction of protein-folding mechanisms from free-energy landscapes
  derived from native structures}.
\newblock Proc. Natl. Acad. Sci. 1999;96:11305--11310.

\bibitem{Klimov00PNAS}
Klimov, D.~K. and Thirumalai, D.
\newblock Native topology determines force-induced unfolding pathways in
  globular proteins.
\newblock Proc. Natl. Acad. Sci. U.S.A. 2000;97:7254--7259.

\bibitem{Ferrara01JMB}
Ferrara, P. and Caflisch, A.
\newblock {Native Topology or Specific Interactions: What is More Important for
  Protein Folding?}
\newblock J. Mol. Biol. 2001;306:837--850.

\bibitem{Swain2006COSB}
Swain, J. and Gierasch, L.
\newblock The changing landscape of protein allostery.
\newblock Curr. Opin. Struct. Biol. 2006;16(1):102--108.

\bibitem{goodey2008NCB}
Goodey, N. and Benkovic, S.
\newblock Allosteric regulation and catalysis emerge via a common route.
\newblock Nature Chem. Biol. 2008;4(8):474--482.

\bibitem{hardy2004COSB}
Hardy, J. and Wells, J.
\newblock Searching for new allosteric sites in enzymes.
\newblock Curr. Opin. Struct. Biol. 2004;14(6):706--715.

\bibitem{Hyeon06PNAS}
Hyeon, C., Lorimer, G.~H., and Thirumalai, D.
\newblock {Dynamics of allosteric transition in GroEL}.
\newblock Proc. Natl. Acad. Sci. U.S.A. 2006;103:18939--18944.

\bibitem{Hyeon11BJ}
Hyeon, C. and Onuchic, J.~N.
\newblock {A Structural Perspective on the Dynamics of Kinesin Motors}.
\newblock Biophys. J. 2011;101:2749--2759.

\end{thebibliography}

\clearpage

\clearpage

{\bf SUPPORTING INFORMATION I}
\\
{\bf Supporting Figures}
\renewcommand{\thefigure}{S\arabic{figure}} 
\setcounter{figure}{0}

\begin{figure}[!ht]
\begin{center}
\includegraphics[width=0.8\columnwidth]{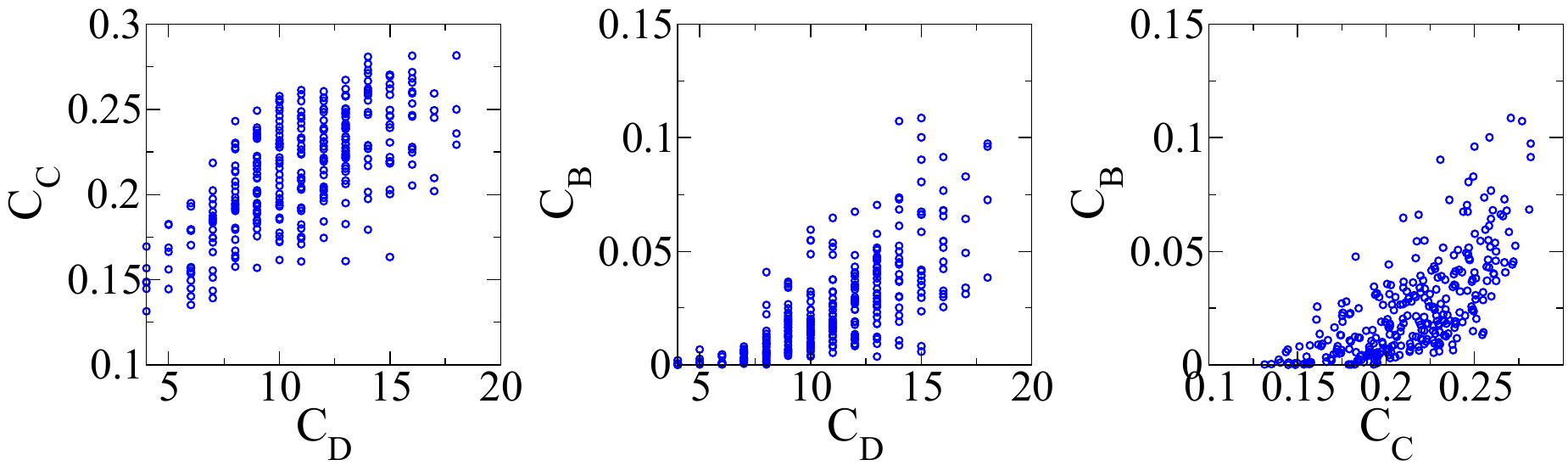}
\end{center}
\caption{ 
Scatter plots of (A) $C_C$ vs $C_D$, (B) $C_B$ vs $C_D$, and (C) $C_C$ vs $C_B$.}
\end{figure}

\begin{figure}[!ht]
\begin{center}
\includegraphics[width=0.8\columnwidth]{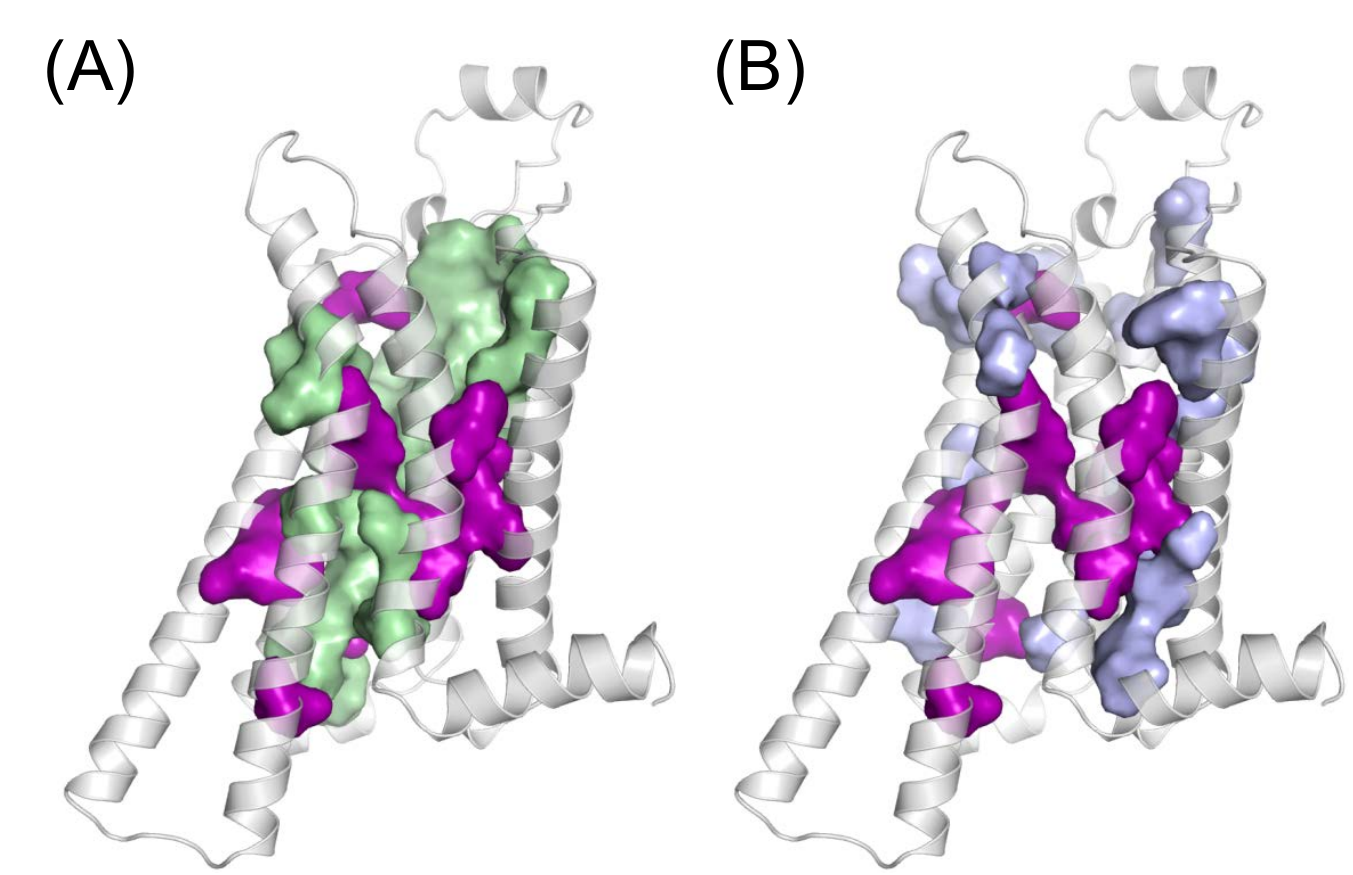}
\end{center}
\caption{ 
Residues with (A) $C_B(\geq 0.05)$ (groups I and II in Figure 2C) and 
               with (B) $\Delta G/k_BT(\geq 1.5)$ in AR family (groups I and III in Figure 2C)
}
\end{figure}

\begin{figure}[!ht]
\begin{center}
\includegraphics[width=0.5\columnwidth]{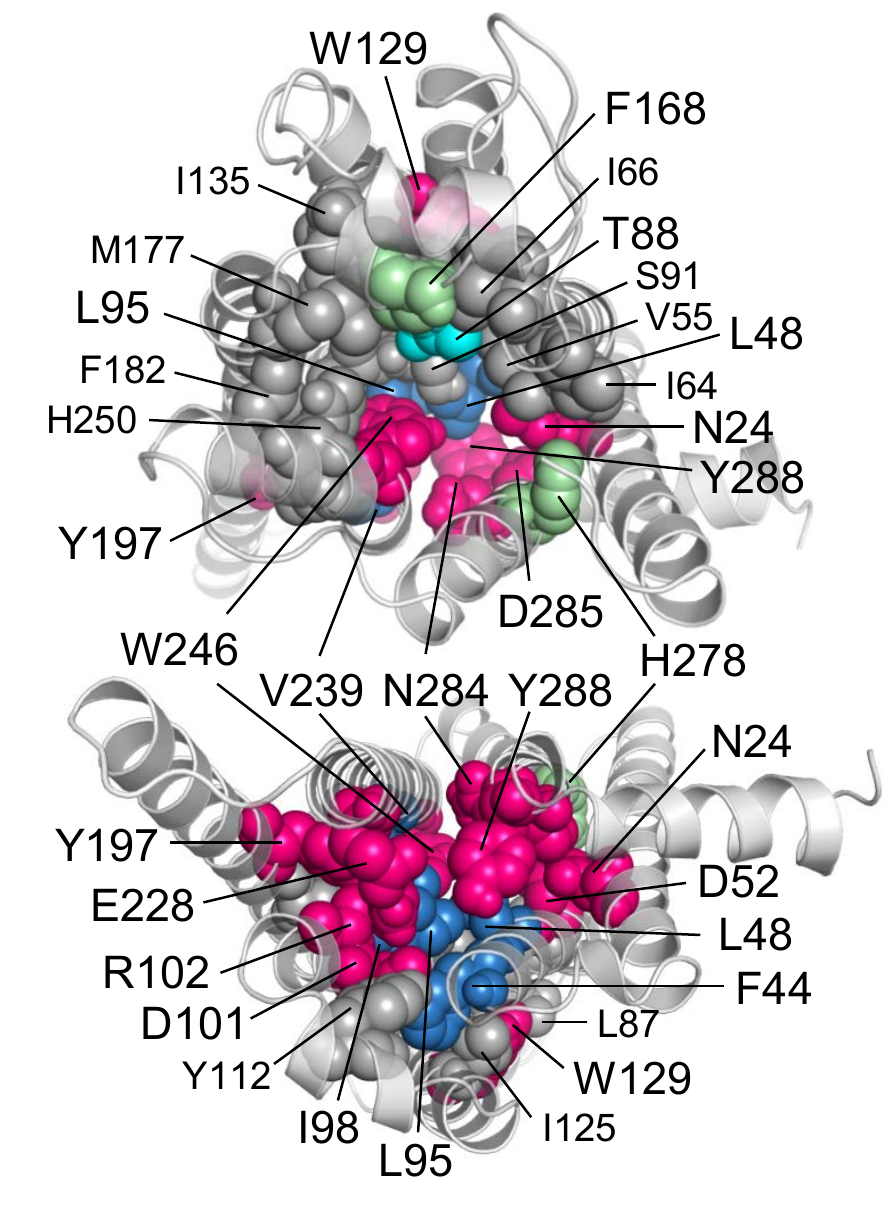}
\end{center}
\caption{ 
The residues with $C_B\geq 0.05$ represented by spheres in the extracellular view (top) and intracellular view (bottom). The color indices are same as in the Figure 2D. 
}
\end{figure}



\begin{figure}[!ht]
\begin{center}
\includegraphics[width=5.7in]{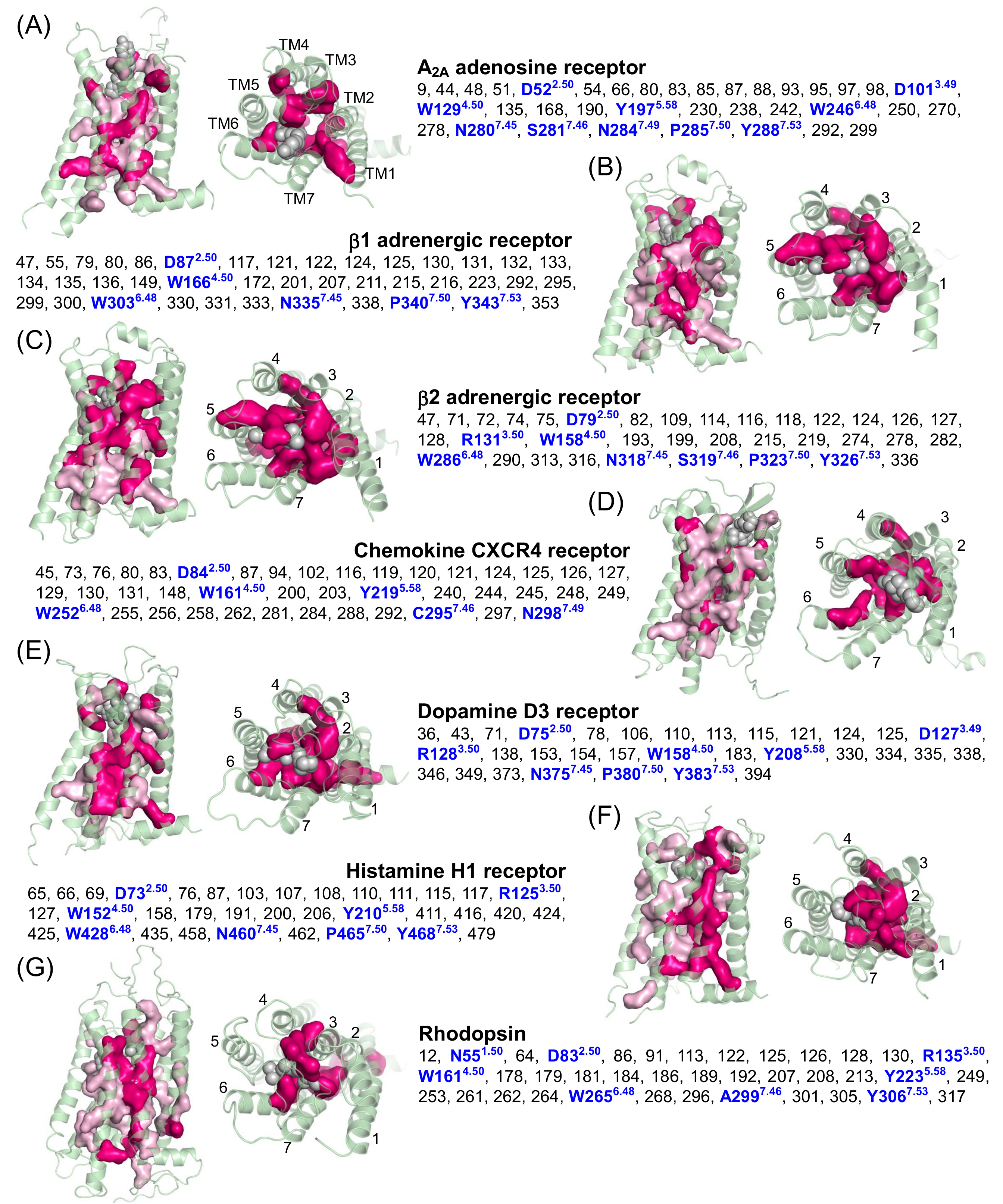}
\end{center}
\caption{
{\bf Network of residues with $C_{B} \geq 0.05$ represented in pink and $C_{B} \geq 0.075$ in magenta for the proteins belonging to the class A GPCR family.}  
The residue networks are depicted using surfaces, and the residue indices are listed. Shown are the side and extracellular (top) views of GPCRs with the bound ligands displayed in gray spheres. The PDB IDs used in the calculations are as follows: (A) A$_{2A}$ adenosine receptor (3EML), (B) $\beta$1 adrenergic receptor (2VT4), (C) $\beta$2 adrenergic receptor (3NYA), (D) Chemokine CXCR4 receptor (3ODU), (E) Dopamine D3 receptor (3PBL), (F) Histamine H1 receptor (3RZE), (G) Rhodopsin (1U19). 
}
\label{Figure_label}
\end{figure}

\begin{figure}
\begin{center}
\includegraphics[width=0.7\columnwidth]{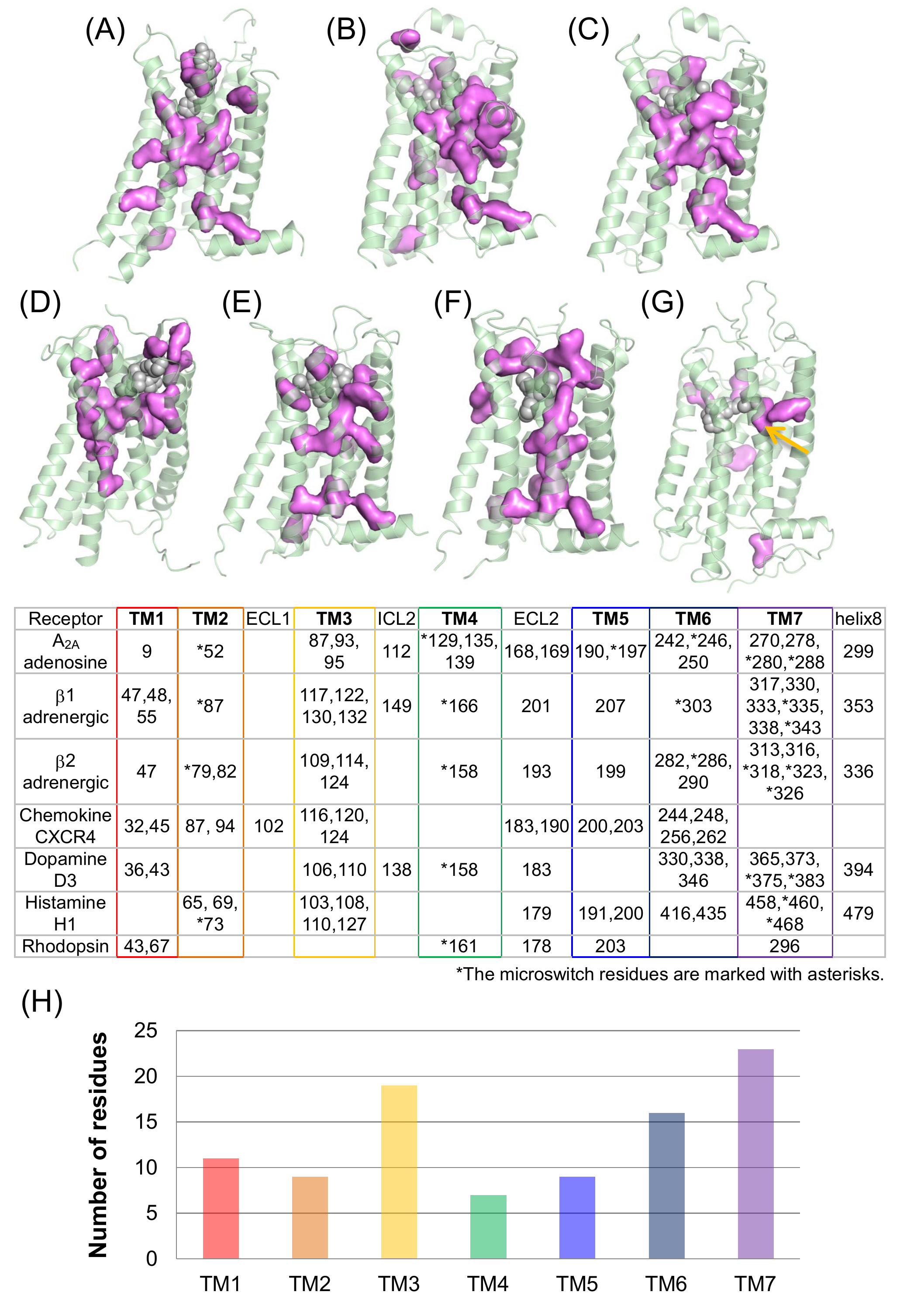}
\end{center}
\caption{
Glycine scanning (network vulnerability) analysis of the crystallized class A GPCRs. (A) A$_{2A}$ adenosine receptor (PDB ID: 3EML). (B) Adrenergic $\beta$1 receptor (2VT4). (C) Adrenergic $\beta$2 receptor (3NYA). (D) Chemokine CXCR4 receptor (3ODU). (E) Dopamine D3 receptor (3PBL). (F) Histamine H1 receptor (3RZE). (G) Rhodopsin (1U19).  The residues with high network vulnerability ($|\Gamma|\geq 0.003$) are depicted using pink surfaces. The bound ligands are shown in gray spheres. The list of residues with $|\Gamma|\geq 0.003$ are given in the table. 
(H) Total number of the highly vulnerable residues in each TM.}
\end{figure}
\clearpage

\begin{figure}
\begin{center}
\includegraphics[width=0.8\columnwidth]{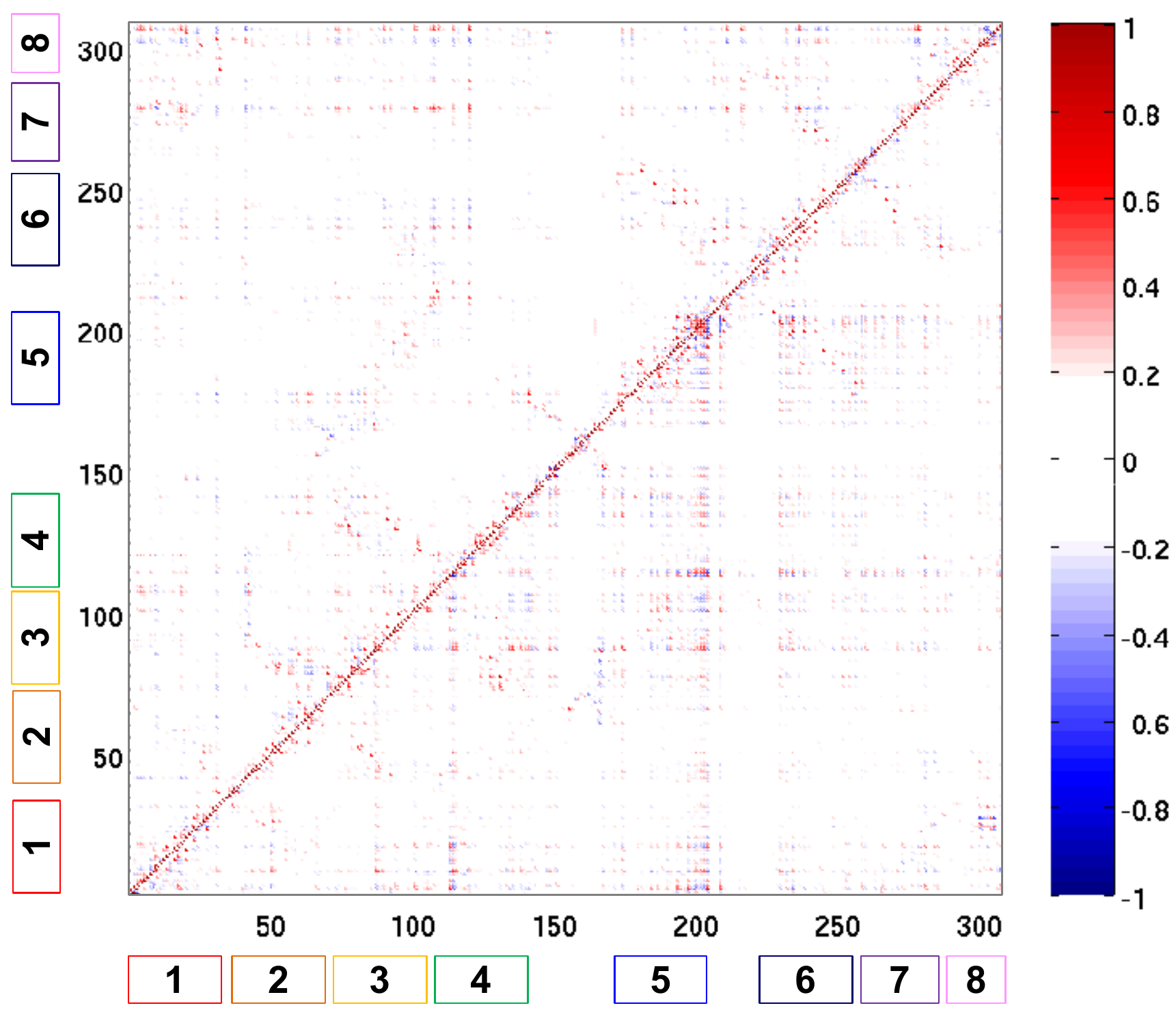}
\end{center}
\caption{
Cross-correlation map of the residue centralities during the 300 nsec MD simulation in the apo form (the upper left panel) and the agonist-bound form (the lower right panel).}
\end{figure}

\begin{figure}[!ht]
\begin{center}
\includegraphics[width=6in]{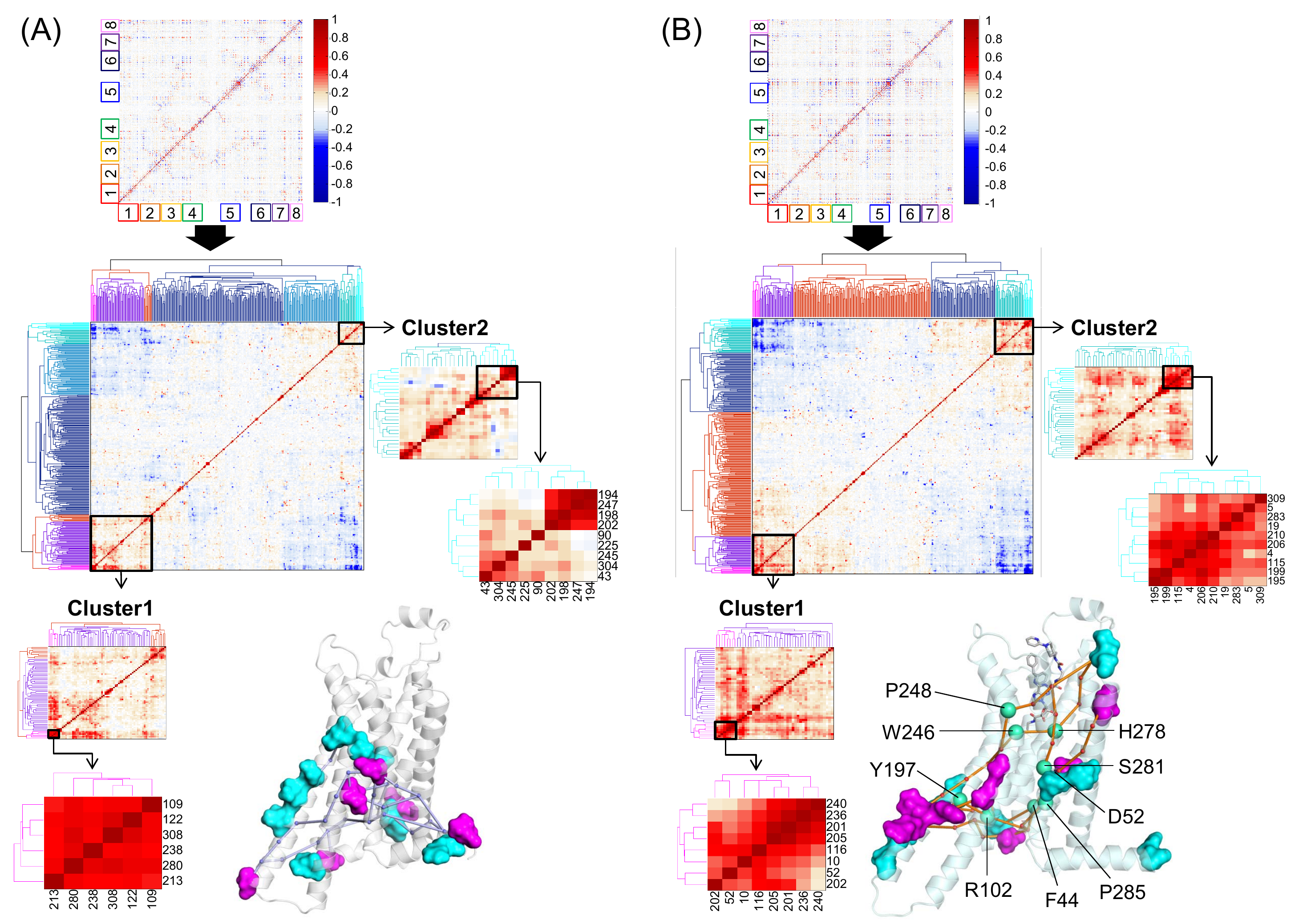}
\end{center}
\caption{
{\bf Clusters of long-range cross-correlated residues calculated for A$_{2A}$AR.} 
(A) Cross-correlation map calculated for the apo form using Eq. 2 and its hierarchical clustering analysis. Highly cross-correlated clusters are enclosed with black boxes (cluster1 and cluster2). In the clusters, cross-correlated residue pairs with $CC_{ij} \geq 0.5$ are represented using cyan and magenta surfaces for the cluster1 and cluster2, respectively. Cross-correlated residue pairs are linked with minimal paths. 
(B) Same calculations for the agonist-bound form. 
}
\label{Figure_label}
\end{figure}

\begin{figure}
\begin{center}
\includegraphics[width=0.9\columnwidth]{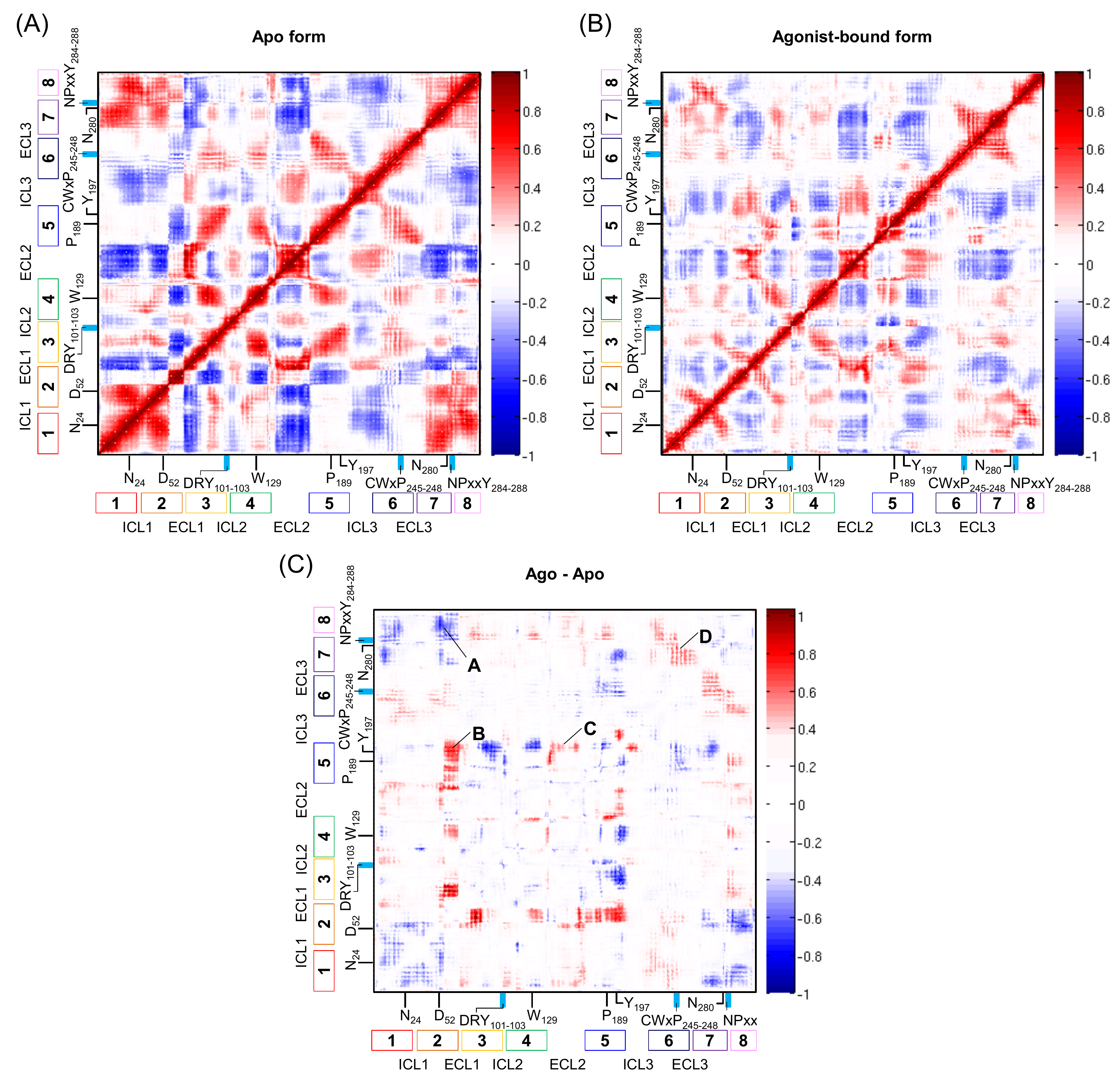}
\end{center}
\caption{Cross-correlations of fluctuations for (A) apo, (B) agonist-bound forms, and (C) their difference calculated with MD trajectories. TM regions (TM1-TM7 and helix 8), intracellular loops (ICLs), extracellular loops (ECLs), and important structural motifs including microswitches are marked. The difference map in (C) shows that there is a dynamic coupling between extracellular ligand binding site and intracellular G-protein binding site: An agonist binding increases the correlation between the extracellular part of TM2 and the intracellular part of TM5 (region marked with ``B'' in (C)); whereas reduces the correlation of the extracellular part of TM2 and TM7-helix 8 (region ``A"). 
In addition, the cross-correlation between the extracellular ligand binding site and intracellular G-protein binding site (``C" and ``D" regions) is increased.   
}
\end{figure}

\begin{figure}
\begin{center}
\includegraphics[width=0.8\columnwidth]{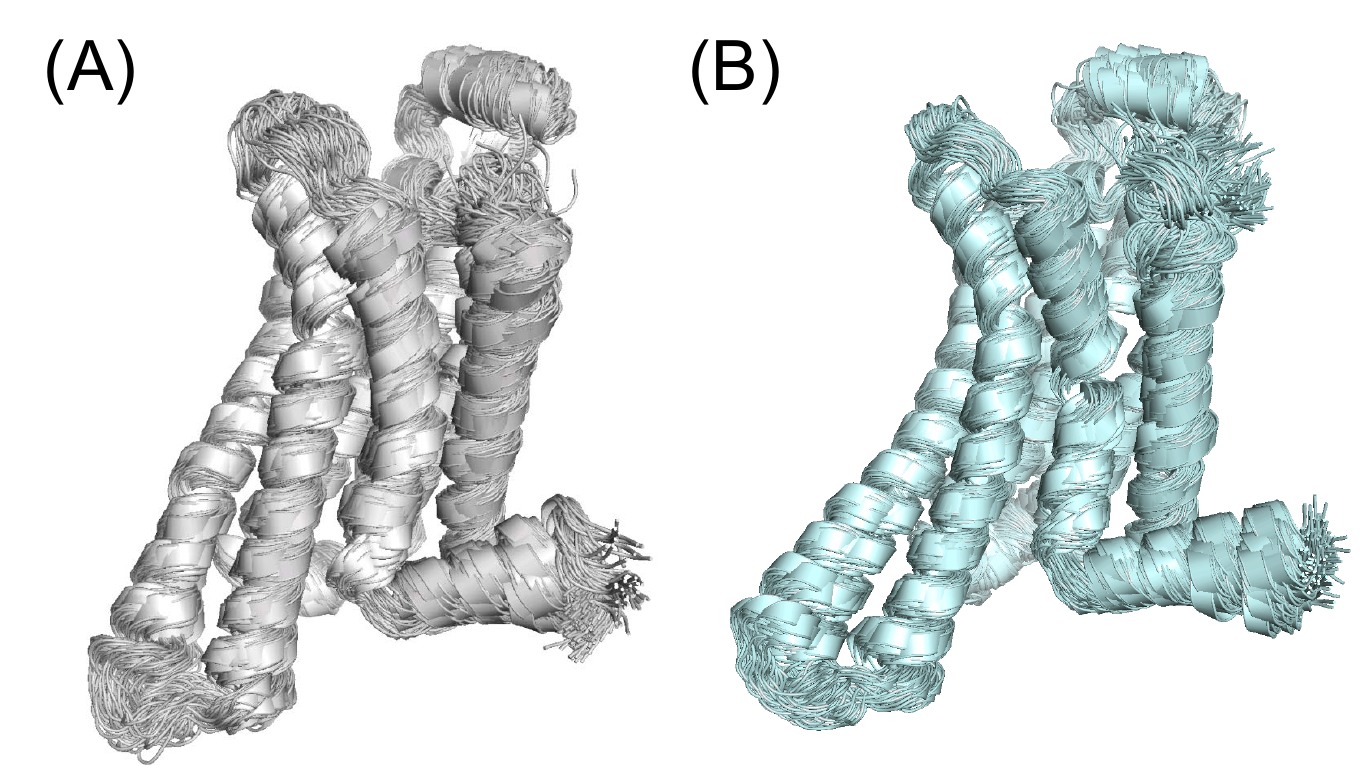}
\end{center}
\caption{Ensemble of structures obtained from MD simulations for the (A) apo and (B) agonist-bound form.   
}
\end{figure}
\clearpage 

{\bf Supporting Information II}

Supporting Information II is downloadable in the following web link: 

\url{http://newton.kias.re.kr/~hyeoncb/homepage/publication/Supporting_Information_II.pdf}



\end{document}